\documentclass{chet}

\usepackage{graphicx}
\usepackage[crop=pdfcrop]{pstool}
\usepackage{float}
\usepackage{dsfont}
\usepackage{mathrsfs}
\usepackage{amsfonts}

\newcommand{\vev}[1]{\langle {#1}\rangle}

\newcommand{\CA}{\mathcal{A}}
\newcommand{\CB}{\mathcal{B}}
\newcommand{\CO}{\mathcal{O}}
\newcommand{\CJ}{\mathcal{J}}
\newcommand{\CT}{\mathcal{T}}

\newcommand{\half}{\frac{1}{2}}
\newcommand{\prim}{\text{prim}}
\newcommand{\osbx}[2]{x_{#1\hspace{1pt}}{\!}^{#2}}

\DeclareMathOperator{\Tr}{Tr}
\DeclareMathOperator{\diag}{diag}

\preprint{UCSD-PTH-11-05}
\title{Current OPEs in Superconformal Theories}
\author{Jean-Fran\c{c}ois Fortin, Kenneth Intriligator and Andreas Stergiou}
\affiliation{Department of Physics, University of California, San Diego, La Jolla, CA 92093 USA}
\abstract{It's well known that in conformal theories the two- and three-point functions of  a subset of the local operators---the conformal primaries---suffice, via the operator product expansion (OPE), to determine all local correlation functions of operators.   It's less well known that, in superconformal theories, the OPE of superdescendants is generally undetermined from those of the superprimaries, and there is no universal notion of superconformal blocks.  We recall these
and related aspects of 4d (S)CFTs, and then we focus on  the super operator product expansion (sOPE) of conserved currents in 4d ${\cal N}=1$ SCFTs.   The current-current OPE $J(x)J(0)$ has applications to general gauge mediation.  We show how the superconformal symmetry, when combined with current conservation, determines the OPE coefficients of superconformal descendants in terms of those of the superconformal primaries.  We show that only integer-spin real superconformal-primary operators of vanishing R-charge, and their descendants, appear in the sOPE.  We also discuss superconformal blocks for four-point functions of the conserved currents.  }

\date{July 2011}

\begin{document}

\maketitle

\newsec{Introduction}

There are many examples of 4d (super)conformal theories ((S)CFTs).  Some have microscopic Lagrangian descriptions, e.g.\ ${\cal N}=1$ SQCD in the conformal window \rcite{Seiberg:1994pq} or ${\cal N}=4$ SYM, while others need not (e.g.\ \rcite{Benini:2009mz}).  Even if there is a microscopic description, it's generally of limited use, because of strong coupling effects.  The ``observables" of conformal theories are the spectrum of operators $\CO _i$, their operator dimensions $\Delta _i$, and their operator product expansion (OPE) coefficients,\foot{There are also non-local observables, like Wilson loops, but we will not discuss them here.} the $c_{ij}^k$ in
\eqn{\CO _i(x)\CO_j(0)=\sum_{\CO_k }\frac{c_{ij}^k}{x^{\Delta _i+\Delta _j-\Delta _k}}\CO _k(0) =\sum _{\substack{\rm{primary}\\ \CO_k}}\frac{c_{ij}^k}{x^{\Delta _i+\Delta _j-\Delta _k}}F_{\Delta _i\Delta _j}^{\Delta _k}(x, P)\CO _k(0).}[opegen]
Conformal symmetry implies that all local operator correlation functions are fully determined, via the OPE, by the $n\leq 3$-point functions of a subset of the operators, the primaries.  In particular, conformal symmetry relates the OPE coefficients of descendant operators to those of the primaries, with determined functions $F_{\Delta _i\Delta _j}^{\Delta _k}(x,P)$ in \opegen.  The OPE expansion \opegen\ is exact in CFTs, and determines all correlation functions of local operators.  We're here interested in 4d ${\cal N}=1$ SCFTs, where the additional symmetry yields additional relations among OPE coefficients.

Conformal or approximately conformal theories are intrinsically interesting, and have various possible applications to high energy physics and beyond the Standard Model (BSM) model building, to perhaps help mitigate various model building challenges.    For example, invoking  running effects with ${\cal O}(1)$ anomalous dimensions could help suppress or enhance otherwise finely tuned quantities or ratios.  Examples include sequestering \rcite{Luty:2001zv}, achieving flavor hierarchy from anarchy \rcite{Nelson:2000sn, Poland:2009yb, Craig:2010ip}, and $\mu / B_\mu$ in gauge meditation\rcite{Roy:2007nz, Murayama:2007ge}.  Furthermore, flowing near an approximate CFT could help lead to useful scale separations or interesting phenomenology, e.g.\ in walking technicolor or unparticles with mass gaps.

Our discussion here is particularly motivated by possible applications to general gauge mediation (GGM)  \rcite{Meade:2008wd}, where one is interested in current-current two-point functions like $\vev{J(x)J(0)}$.  4d ${\cal N}=1$ supersymmetry conserved currents $j_\mu$ reside in real supermultiplets
\eqn{\CJ(z)=J(x)+i\theta j(x)-i\bar\theta \bar\jmath(x) -\theta \sigma ^\mu \bar \theta j_\mu(x) +\cdots,}[Jzis]
where $\cdots$ are derivative terms, following from the conservation equations $D^2\CJ =\bar D^2 \CJ =0$.   The operator\footnote{We use $|$ to denote the bottom component, setting all $\theta , \bar \theta =0$.}  $J(x)=\CJ |$ is a real superconformal primary, with dimension $\Delta _J=2$, and the conserved current $j_\mu (x)$ is among its descendants.   Here $j_\mu (x)$ is a global current of the CFT (that could later be weakly gauged as in GGM).
With this application in mind, we will here consider general aspects of the super OPEs of these operators in 4d ${\cal N}=1$ SCFTs.  We will discuss  applications to GGM in detail in a separate paper \rcite{Fortin:2011aa}.

The leading short-distance terms in the OPE of $\CJ (z)$ with operators have universal coefficients, fixed in terms of the charges.   As we'll recall, this is similar to the universal coefficients in OPEs involving the conserved  $\CT _\mu(z)$ $U(1)_R$-plus-stress-energy-tensor supermultiplet\rcite{Ferrara:1974pz} of SCFTs, which was considered e.g.\ in \rcite{Anselmi:1996dd, Anselmi:1997am, Osborn:1998qu}.  The leading terms in the OPE of the bottom, primary component of currents with themselves take the form
\eqn{J_a(x) J_b(0)=\tau \frac{\delta _{ab}\mathds{1}}{16\pi ^4x^4}+\frac{k
d_{abc}}{\tau} \frac{J_c(0)}{16\pi ^2x^2}-f_{abc} \frac{x^\mu j_\mu ^c
(0)}{ 24\pi ^2 x^2}+c_{ab}^i\frac{\CO_i(0)}{x^{4-\Delta _i}}+\cdots,}[jjex]
with $a$ an adjoint index for the (say simple) group $G$. In what follows, we often suppress the group adjoint index, or simply take $G=U(1)$ since the generalization is fairly straightforward.  For the moment, we just want to illustrate a point with the symmetric $d_{abc}$ and the structure function terms $f_{abc}$ in \jjex.

Conformal symmetry relates terms in the OPE.    In the non-SUSY case, the coefficients of all descendant operators are fully determined from those of the primary operators, as was worked out
(in many different ways) in the 1970s, see e.g.\  \rcite{Ferrara:1973yt}.  It is natural to expect that (i) the SUSY version should be completely analogous and (ii) that it must have long ago been worked out for general operators.  But both statements are untrue!  This follows from the works of Hugh Osborn and collaborators, but it has not been very explicitly discussed in the literature, and it comes as an initial surprise to many experts.

The OPE is  related to operator two- and three-point functions, and the fact that non-SUSY conformal descendant terms are uniquely characterized by the primaries is related to the fact that conformal symmetry can be used to map any three operator-insertion points $x_{1,2,3}^\mu$ to wherever one pleases.  The constraints of  (non-SUSY) conformal symmetry on operator two- and three-point functions, in general spacetime dimension $d$, were studied in \rcite{Osborn:1993cr}, including the additional constraints coming from Ward identities for conserved quantities like $j_\mu$ or $T_{\mu \nu}$.

That the OPE coefficients of superconformal primaries are generally {\it not} sufficient to determine those of the superdescendants can likewise be understood from their relation to operator two- and three-point functions.  The 4d ${\cal N}=1$ superconformal constraints on operator two- and three-point functions were analyzed, using a superspace analysis by Osborn \rcite{Osborn:1998qu}, and we'll here review, and heavily use, his framework.  A quick way to understand why superdescendant three-point functions are generally not fully determined by the primaries is to note that ${\cal N}=1$ supertranslations and superconformal transformations  only suffice to eliminate the Grassmann coordinates at two points in superspace---the third Grassmann coordinate in three-point functions remains.  This explains the existence of the nilpotent three-point function superconformal invariant building blocks, $\Theta$ and $\bar \Theta$, found in superspace in \rcite{Osborn:1998qu} (see also \rcite{Park:1997bq}).

 As an illustration, consider the superspace expression for current three-point functions \rcite{Osborn:1998qu}, capturing the $G$ structure functions $f_{abc}$ and $\Tr G^3$ 't Hooft anomaly $k$,
\eqn{\vev{\CJ _a(z_1)\CJ _b(z_2) \CJ _c(z_3)}=\frac{1}{\osbx{\bar 31}{2}\osbx{\bar 13}{2} \osbx{\bar 32}{2}\osbx{\bar 23}{2}}\left[i\frac{f_{abc}\tau}{128\pi ^6}\left(\frac{1}{X_3^2}-\frac{1}{\bar X _3^2}\right)+\frac{d_{abc}k}{256\pi ^6}
\left(\frac{1}{X_3^2}+\frac{1}{\bar X_3^2} \right)\right],}[JJJsope]
with notation reviewed in section \ref{Osb}.  For now we will just say that $X-\bar X=4i\Theta\bar\Theta$, with $\Theta \sim \theta$'s in superspace.  The $f_{abc}$ terms in \JJJsope\ do not contribute if we restrict (via $\theta \to 0$) to superconformal primary components, but do contribute for  superdescendants.   Explicitly, in \jjex, the $f_{abc}$ term is a descendant coefficient that is unrelated to the $kd_{abc}$ primary coefficient.
In \JJJsope\ the $\Theta $ dependence is at least determined by $G$ symmetry.  For general operators, the $\Theta$ dependence is ambiguous, not fully determined by the symmetries.

We will here study the general constraints of superconformal symmetry on
the  two- and three-point functions relevant for the $J(x)J(0)$ sOPE, and
how the sOPE coefficients are obtained from these correlators.  We will do
this both using the  superspace results of Osborn \rcite{Osborn:1998qu} for
the relevant two- and three-point functions, and also directly from the
superconformal algebra.  As we'll discuss, the fact that the currents are
conserved here allows the superspace $\Theta$ dependence to be completely
fixed.  Thus, the coefficients of the superconformal primaries in the
$J(x)J(0)$ OPE suffice to fully determine all OPE coefficients of all
descendants.    We will discuss the contributions on the RHS of the
$J(x)J(0)$ OPE from integer-spin real $U(1)_R$-charge-zero superconformal
primaries, $\CO ^{\mu _1\dots \mu _\ell}$, and their
superdescendants.\foot{Note added (April 2014 revision): as was later found
in~\cite{Berkooz:2014yda}, there are additional contributing Lorentz
representations.   This revised version will also correct a couple of
errors in our original version's coefficients, as pointed out to us by the
authors of~\cite{Berkooz:2014yda} and \cite{Khandker:2014mpa}; see these
papers for further details. We thank the authors of \cite{Berkooz:2014yda,
Khandker:2014mpa} for bringing these issues to our attention, and for
sharing their interesting results with us prior to their general posting.}

The paper is organized as follows: section \ref{OPE} briefly reviews the aspects of the OPE in 4d CFTs that we will use in the following discussion.  Section \ref{scOPE} discusses superconformal theories, and the constraints of superconformal symmetry on two- and three-point functions and the OPE.  The superspace formalism of \rcite{Osborn:1998qu}, and the recent results about chiral-chiral and chiral-anti-chiral OPEs \rcite{Dolan:2000uw,Poland:2010wg,Vichi:2011ux}, are reviewed.   In section \ref{JJOPEsec} we consider the current-current OPE, showing how the additional constraints of the current's conservation constrains the $\vev{JJ\CO}$ three-point functions, and hence the OPE.  We show that only real, $U(1)_R$-charge zero, integer-spin operators $\CO ^{(\ell)}$, and their superconformal descendants, can appear on the RHS of the $J(x) J(0)$ OPE.  We show that the OPE coefficients within each supermultiplet are fully specified by a single OPE coefficient.   The dependence on the nilpotent invariant $\Theta$ mentioned above is here fully determined by the ${\cal J}$ current conservation.

In section \ref{JJJJsec} we discuss aspects of four-point functions and their conformal blocks, where the four-point function is factorized into an OPE sum of intermediate operators, and their descendants, in the $s$, $t$, or $u$ channel.   In ${\cal N}=0$ theories, the contribution of an intermediate primary operator of dimension $\Delta$ and spin $\ell$ is given by a known function \rcite{Dolan:2000ut}, $g_{\Delta, \ell}(u, v)$, which accounts for the sum over descendants and is independent of the external operators.  There is no general analog of such a general ``superconformal block" in SCFTs, because of the generally ambiguous dependence on the super-descendants in the sOPE.  This ambiguity is resolved when the external operators are in reduced multiplets, in particular the chiral and anti-chiral multiplets discussed in \rcite{Poland:2010wg} and the conserved currents discussed here.  The superconfomal blocks, then, depend on the type of external states.  We review the results of \rcite{Poland:2010wg} for ${\cal N}=1$ superconformal blocks ${\cal G}_{\Delta, \ell}^{\phi \phi ^*; \phi \phi ^*}$, and briefly mention how ${\cal G}_{\Delta, \ell}^{\phi \phi; \phi ^*\phi ^*}$ differs.  Then we discuss the ${\cal N}=1$ superconformal blocks for ${\cal G}_{\Delta \ell}^{JJ; JJ}$ and ${\cal G}_{\Delta, \ell}^{JJ; \phi \phi ^*}$.  Finally, we discuss these quantities in ${\cal N}=2$ SCFTs, where they are related by the additional $SU(2)_I$ symmetry.

Section \ref{CONC} summarizes our findings and discusses possible applications of the results.   Finally, appendix \ref{SCA} summarizes some of the relations of the (super)conformal algebra, and our sign conventions.


\newsec{Review of OPE results in the non-SUSY case}[OPE]

Aspects of CFTs and the OPE are discussed in many references and reviews.  We will here review, for completeness,  some of the main points for our later use.   We summarize the algebra and our sign conventions in appendix \ref{SCA}.

\subsec{Primaries and descendants and their two- and three-point functions}

Representations of the conformal group are built by regarding $P_\mu$ and $K_\mu$ as raising and lowering operators, respectively; they raise or lower operator dimension by one unit.  Each irreducible representation has a  lowest, ``quasi-primary" operator at the bottom, which is annihilated by all lowering operators at the origin,  $x^\mu =0$. (The origin is a distinguished point, as the fixed point of scale transformations.)  The quasi-primary has an associated tower of ``descendant" operators above it, generated by $[P_\mu , \star]$; this accounts for the fact that the
operators can anyway be translated to a general point via $\CO^I(x)=e^{-iP\cdot x}\CO ^I(0)e^{iP\cdot x}$.

Conformal symmetry completely determines the form of the $n\leq 3$-point functions,  in terms of the operator dimensions, up to the overall normalization coefficients.  This follows from the fact that conformal transformations can be used to map any three points $x^\mu _{1,2,3}$ to wherever one pleases.  For example, we can use translation symmetry to map $x^\mu _1=0$, and special conformal symmetry to make $x^\mu _3=\infty$, and then use Lorentz symmetry and dilatations to map $x_2^\mu$ to a canonical unit vector.

Scale invariance implies that the only non-zero one-point function is that of the identity operator,
which is the only operator with $\Delta _\CO =0$:
\eqn{\vev{\CO _a (x)}=\delta _{a,0}, \qquad  \CO _0\equiv \mathds{1}.}[onepoint]

The two-point functions of primary operators take the form
\eqn{\vev{\CO _i^{s_i}(x_i)\CO _j^{s_j}(x_j)}=\frac{c_{ij}}{r_{ij}^{\Delta _i}}P^{s_is_j}(x_{ij}), \qquad x_{ij}^\mu \equiv x_i^\mu -x_j^\mu , \qquad r_{ij}\equiv x_{ij}^2.}[twopoint]
Here $c_{ij}$ are constant normalization coefficients, the analog of the Zamolodchikov metric on the space of deformations in 2d.  Conformal symmetry implies that $c_{ij}$ vanish unless the two operators have the same operator dimension, $c_{ij}\propto \delta _{\Delta _i, \Delta _j}$, and of course the same spin.  The $s_{i,j}$ in \twopoint\ are Lorentz indices and $P^{s_is_j}(x)$ is an appropriate representation of the rotation group, e.g.\ $P=1$ for scalars or,  taking both operators to have spin $\ell$, with $s_i=(\mu _1\dots \mu _\ell)$ and  $s_j=(\nu _1\dots \nu _\ell)$, both symmetrized and traceless \rcite{Osborn:1993cr},
\eqn{P^{s_is_j}(x)=I^{(\mu _1 \nu _1}(x)\cdots I^{\mu _\ell \nu _\ell)}(x), \qquad I^{\mu \nu}(x)\equiv \eta ^{\mu \nu}-2\frac{x^\mu x^\nu}{x^2},}[OOspin]
with the Lorentz indices symmetrized and traceless.

Conformal symmetry implies that primary operator three-point functions have the form
\eqn{\vev{\CO _i^{s_i}(x_i)\CO _j^{s_j}(x_j)\CO _k^{s_k}(x_k)}=\frac{c_{ijk}}{r_{ij}^{\frac12(\Delta _i+\Delta _j-\Delta _k)}r_{ik}^{\frac12(\Delta _i+\Delta _k-\Delta _j)}r_{jk}^{\frac12(\Delta _j+\Delta _k-\Delta _i)}}P^{s_is_js_k}(x),}[threepoint]
where $c_{ijk}$ are constants and $P^{s_is_js_k}(x)$ is a fixed tensor depending on the Lorentz spins of the operators, e.g.\ $P=1$ for scalar operators, that is determined in \rcite{Osborn:1993cr}.
Of course, \threepoint\ reduces to \twopoint\ if any of the operators is the identity, so $c_{0ij}=c_{ij}$.
 A case of particular interest here is for two scalar primaries and one spin-$\ell$ primary operator, where the explicit form of \threepoint\ is
\eqn{\vev{\CO _i(x_i)\CO _j (x_j)\CO _k^{(\mu _1\dots \mu _\ell)}(x_k)}=\frac{c_{ijk}}{r_{ij}^{\half (\Delta _i+\Delta _j-\Delta _k+\ell)}r_{ik}^{\half (\Delta _k+\Delta _{ij}-\ell)}r_{jk}^{\half (\Delta _k-\Delta _{ij}-\ell)}}Z^{(\mu _1}Z^{\mu _2}\cdots Z^{\mu _\ell)},}[OOOspin]
where  $\Delta _{ij}\equiv \Delta _i-\Delta _j$, and
\eqn{Z^\mu \equiv \frac{x_{ki}^\mu}{r_{ik}}-\frac{x_{kj}^\mu}{r_{jk}}, \qquad Z^2=\frac{r_{ij}}{r_{ik}r_{jk}},}[Zis]
which is called $X^\mu _{ji}$ in the notation of \rcite{Osborn:1993cr} and $X_k^\mu |_{\theta, \bar \theta =0}$ in the notation of \rcite{Osborn:1998qu} that we'll use shortly.

The primary two- and three-point functions \twopoint\ and \threepoint\ fully determine those of all descendants.   For example, we can replace $\CO _j^{s_j}(x_j)$ with $[P_\mu , \CO _j^{s_j}(x_j)] = i\partial _\mu  \CO _j^{s_j}(x_j)$ in \twopoint\ and \threepoint  simply by taking $i\partial / \partial x_j^\mu$ of the LHS.

The above expressions can be written in terms of (radial quantization) states: using translation symmetry to map $x_i\to 0$, the (say scalar) operator $\CO _i(x_i)$ creates an in-state,
\eqn{\lim _{x_i\to 0}\CO _i (x_i)|0\rangle = |\CO _i\rangle.}[instate]
Using conformal symmetry to map $x_j\to \infty$, $\CO_j(x_i)$ likewise creates an out-state,
\eqn{\lim _{x_j\to \infty}\langle 0|\CO _j(x_j) x_j^{2\Delta _j}=\langle \CO _j|,}[outstate]
where the $x_j^{2\Delta _j}$ factor follows, for example, via an inversion, $x_{\mu}'=x_\mu /x^2$, with $\CO _j'(x')=\Omega^{\rm inv}(x)^{\Delta _j}\CO _j(x)$, $\Omega ^{\text{inv}}(x)=x^2$ (see appendix \ref{SCA}),
which maps \instate\ to  \outstate.  Then, \twopoint\ and \OOOspin\ give (taking $(x_i, x_j, x_k)\to (0, \infty, x)$, \Zis\ gives $Z^\mu \to x^\mu /x^2$)
\eqn{\langle \CO _j|\CO _i\rangle = c_{ij},}
\eqn{\langle \CO _j| \CO _k^{(\mu _1\dots \mu _\ell)}(x)|\CO _i\rangle=  \frac{c_{ijk}}{(x^2)^{\half (\Delta _i+\Delta _j -\Delta _k+\ell)}}x^{(\mu _1}\cdots x^{\mu _\ell)}.}

\subsec{The OPE; descendants from primaries}

The OPE contains precisely the same information as the two- and three-point functions:
\eqn{\CO _i ^{s_i}(x_i)\CO _j ^{s_j}(x_j)=\frac{c_{ij} P^{s_is_j}(x_{ij})}{r_{ij}^{\Delta _i}} \mathds{1} +\sum _{k'} \frac{c_{ij}^{k'}}{r_{ij}^{\half (\Delta _i+\Delta _j -\Delta _{k'})} }  [F_{ij}^{k'} (x_{ij}, P),\CO _{k'} ]^{(s_{k'})}(x_j).}[opeeg]
The function $F_{ij}^{k'} (x_{ij}, P)$ gives the coefficients of the descendant operators and depends only on the operator dimensions $\Delta _{i, j, k'}$ and spins $s_{i, j, k'}$.  Taking expectation values of both sides yields \twopoint\ from the unit operator $\CO _0\equiv \mathds{1}$ on the RHS of \opeeg, so $c_{ij}=c_{ij}^0$.

To relate the OPE \opeeg\ to the three-point functions \threepoint we multiply both sides of \opeeg\ by $\CO _{k}^{s_k}(x_k)$ and then, taking the expectation value, use \twopoint\ to evaluate the remaining two-point function
$\vev{\CO _{k'} ^{s_{k'}}(x_j)\CO _k ^{s_k}(x_k)}$.   This gives the relation
\eqn{c_{ijk}=c_{ij}^{k'}c_{kk'},\qquad  \text{or equivalently}\qquad  c_{ij}^k=c_{ijk'}c^{k'k} \qquad\text{for primaries},}[craise]
where $c^{kk'}c_{k'm}=\delta ^k_m$, summing the dummy index $k'$.   It follows from \craise\ that, e.g.\ \eqn{c^k_{ij}=c^{k\ell}c_{jm}c_{i\ell}^m.}[ccreln]

The relations \craise\ follow from matching the OPE \opeeg\ to merely the leading $x_{ij}\to 0$ dependence in the three-point functions \threepoint.    This leading dependence comes from restricting to primary operators on the RHS of the OPE, dropping the $[P_\mu, \star]$ descendant terms.    Matching to the full $x_{ij}$, $x_{jk}$, and $x_{ik}$ dependence in \threepoint\ will determine the coefficients of all the $[P_\mu, \star]$ descendant terms, i.e.\ the function $F_{ij}^k (x_{ij}, P)$, in the OPE \opeeg.     These functions incorporate also the spin dependence, which is a complication that we won't need to deal with in full generality.  It'll suffice here to focus on the OPE of scalar operators.

Consider then the OPE of two scalar operators, which generally includes non-zero integer-spin-$\ell$ primary operators  $\CO _{k'}^{(\mu _1\dots \mu _\ell)}$ (with symmetrized indices) on the RHS,
\eqn{\CO_i(x_i)\CO _j(x_j)=\sum _{\CO _{k'}^{\ell}}\frac{c_{ij}^{k'}}{r_{ij}^{\half (\Delta _i+\Delta _j-\Delta _{k'})}}F_{\Delta _i\Delta _j}^{\Delta _{k'}; \ell }(x_{ij}, P)_{\mu _1\dots \mu _\ell}\CO _{k'}^{(\mu _1\dots \mu _\ell)}(x_j).}[OPEspin]
The (odd) even spin $\ell$ terms are (anti-) symmetric under $\CO _i\leftrightarrow \CO _j$.
For simplicity, consider first the spin $\ell =0$ primary operators on the RHS,
\eqn{\CO _i (x_i)\CO _j(x_j) \supset \sum _{\CO_{k'}^{\ell =0}} \frac{c_{ij}^{k'} }{r_{ij}^{\frac{1}{2}(\Delta _i +\Delta _j-\Delta _{k'} )}}F_{\Delta _i\Delta _j}^{\Delta _{k'} }(x_{ij},P)\CO _{k'}(x_j).}[primOPE]
The function $F_{\Delta _i\Delta _j}^{\Delta _{k'} }(x,P)$ satisfies $F_{\Delta _i\Delta _j}^{\Delta _{k'}}(x=0,P)=1$, to give the leading $x_{ij}\to 0$ singularity from the primary $\CO _{k'}$.  The higher-order terms in $F$ account for the OPE coefficients of $\CO _{k'}$'s descendants, which are fully determined by the conformal symmetry; reproducing the three-point functions gives one derivation  \rcite{Ferrara:1973yt}:   we multiply \primOPE\ by $\CO _k(x_k)$ and take expectation values of the resulting two-point function using \twopoint, with $P=1$ for this scalar case, and then require that the result reproduces the three-point functions  \threepoint, again with $P=1$.  This determines that,  for this scalar case,
\eqn{F_{\Delta _i\Delta _j}^{\Delta _k}(x_{ij}, P\to i\partial _{x_j})=C^{\half (\Delta _k+\Delta _i-\Delta _j), \half (\Delta _k-\Delta _i+\Delta _j)}(x_{ij}, \partial _{x_j}),}[FCreln] where the function on the RHS is defined to be the solution of
\eqn{C^{ab}(x_{ij}, \partial _{x_j})\frac{1}{r_{jk}^{a+b}}=\frac{1}{r_{ik}^ar_{jk}^b},}[Ceqn]
(see e.g.\ \rcite{Dolan:2000uw}\ for details, as well as the generalization for the general spin-$\ell$ operators) such that \OPEspin\ reproduces the three-point functions \OOOspin.

One can also obtain the functions  $F_{\Delta _i\Delta _j}^{\Delta _k}(x,P)$ that capture the descendant OPE coefficients by requiring that  $[K_\mu, \star]$ gives the same result when taking $\star=$ the LHS and the RHS of  \primOPE.  Using the algebra and action of $K_\mu$, given in appendix \ref{SCA}, this gives
\eqn{i(x^2\partial _\mu -2x_\mu x\cdot \partial -2\Delta _ix_\mu)\left(\frac{F_{\Delta _i\Delta _j}^{\Delta _k}(x,P)}{(x^2)^{\frac{1}{2}(\Delta _i +\Delta _j-\Delta _k)}}\right)=\frac{1}{(x^2)^{\frac{1}{2}(\Delta _i +\Delta _j-\Delta _k)}}[K_\mu, F_{\Delta _i\Delta _j}^{\Delta _k}(x,P)],}[Kmu]
treating the primaries $\CO _k$ as a basis of independent operators.  This equation can be solved exactly, see the original papers \rcite{Ferrara:1973yt}.  As an expansion in powers of $x$, it is straightforward to use the algebra to see that \Kmu\ is solved by
\eqn{F_{\Delta _i\Delta _j}^{\Delta _k}(x,P)=1-\frac{i}{2}\left(\frac{\Delta _k+\Delta _i-\Delta _j}{\Delta _k}\right) x\cdot P+\cdots.}

\subsec{Conserved-current leading OPE singularities from their charges}[ccsings]

The normalization of conserved currents, their leading OPE with other operators and themselves, is determined in terms of the operator's conserved-charge value.    Conserved currents $j_\mu ^a(x)$ are real, spin-$\ell =1$, $\Delta _{j^\mu }=3$ operators.  For simplicity, consider first the case of a $U(1)$ current, $j_\mu (x)$, in the three-point function with a scalar operator of $U(1)$ charge $q_\CO$,
\eqn{\vev{\CO (x_1)\CO ^\dagger (x_2) j^\mu (x_3)}=  -iq_\CO \frac{c_{\CO \CO ^\dagger}}{2\pi ^2}\frac{Z^\mu}{r_{12}^{\Delta _\CO -1}r_{13} r_{23}},}[OOJi]
where we use \OOOspin.  The $i$ is needed for $j^\mu$ to assign the correct charge to the operator, and it ensures that \OOJi\ is Hermitian with the exchange $x_1\leftrightarrow x_2$, which takes $Z^\mu \to -Z^\mu$.  More generally, the OPE of a conserved current $j^a_\mu (x)$ with primary operator $\CO_I(x)$ ($a$ is an adjoint index and $I$ runs over $\CO$'s representation) is
\eqn{j^a_\mu (x)\CO _I(0)=-i(t^{a}_{\CO})_{IJ}\frac{x_\mu}{2\pi ^2x^4}\CO _J(0)+\hbox{less singular},}[JOope]
where $t_{\CO}^{a}$ is the representation of the operator; for a $U(1)$ current, $t_{\CO}=q_\CO$ the $U(1)$ charge, and we take $\CO$ to be a Lorentz scalar for simplicity. For an operator $J^b$ in the adjoint representation, $(t^a)_{bc}=if_{abc}$ so \JOope\ becomes
\eqn{j^a_\mu (x)J^b(0)=f^{abc} \frac{x_\mu}{2\pi ^2x^4}J_c(0)+\hbox{less singular}.}[Jadjope]
Using \ccreln\ with \JOope\ determines the coefficient of $j^a_\mu$ on the RHS of the $\CO^\dagger _I(x)\CO _J(0)$ OPE.  In particular, \Jadjope\ leads to the $f_{abc}$ term on the RHS of \jjex.

The OPE of the stress-energy tensor with the operator is \rcite{Osborn:1993cr}
\eqn{T_{\mu \nu}(x)\CO(0)=-2 {\Delta _\CO}\frac{x^\mu x^\nu -\frac{1}{4}\eta _{\mu \nu}x^2}{3\pi ^2 x^6}\CO(0)+\text{less singular}.}[TOope]
It follows from \OOJi\ and \TOope\ and \ccreln\ that (using $c_{TT}=40c/\pi ^4$)
\eqn{\CO ^\dagger (x)\CO (0)=\frac{c_{\CO ^\dagger \CO}}{x^{2\Delta _\CO}}\mathds{1}-iq _\CO \frac{2\pi ^2}{3\tau} \frac{c_{\CO ^\dagger\CO}x_\mu}{x^{2(\Delta _\CO -1)}}j^\mu (0) +\Delta _\CO \frac{\pi ^2}{60c} \frac{c_{\CO ^\dagger \CO}x_\mu x_\nu}{x^{2(\Delta _\CO -1)}}T_{\mu \nu}(0)+\dots }[OOjT]

These relations between the leading singularities and the charges can be shown, much as in 2d, by computing the charge operator by  integrating the current over a spatial $S^3$ in radial quantization, and then using the OPE where it hits the other operators.
Properly regulated, this yields the commutator of the charge with the operator and the leading singularity gives the operator's charge value.   Equivalently, the leading term coefficients in  \JOope\ and \TOope\ are fixed as they give the correct contact terms in the conserved current's Ward identities for $\partial ^\mu j_\mu$, $\partial ^\mu T_{\mu \nu}$, and $T^{\phantom{\mu}\!\mu} _\mu$. This can be shown \rcite{Osborn:1993cr} by treating the $x\to 0$ singularities in \JOope\ and \TOope\ with differential regularization \rcite{Freedman:1991tk}:
\eqn{{\mathscr{R}}\left(\frac{1}{ x^{2\eta }}\right)=\frac{1}{x^{2\eta}}-\frac{\mu ^{2\eta -4}}{4-2\eta}2\pi ^2 \delta ^{(4)}(x)=-\frac{1}{4-2\eta}\partial ^2 \left(\frac{1}{2\eta -2}\frac{1}{x^{2\eta -2}}-\frac{\mu ^{2\eta -4}}{2}\frac{1}{x^2}\right),}
and for $2\eta \to 4$,
\eqn{{\mathscr{R}}\left(\frac{1}{ x^{4}}\right)=-\frac{1}{4}\partial^2\left(\frac{1}{x^2}\ln(\mu^2x^2)\right).}

The normalization of the currents is fixed by the above conditions, that their OPEs with operators give the correct operator charges.  The  leading singularities in the self-OPEs $j^a_\mu (x)j^b_\nu (0)$ and $T_{\mu \nu}(x)T_{\rho \sigma}(0)$ are similarly determined from Ward identity contact terms.
 The current-current OPE leading terms are
\eqn{j^a_\mu (x)j^b_\nu (0)=3\tau ^{ab}\frac{I_{\mu \nu}(x)}{4\pi ^4x^6}\mathds{1}+2f^{ab}_c\frac{x_\mu x_\nu x^\kappa }{\pi ^2x^6}j^c_\kappa (0) +kd^{ab}_c\frac{D_{\mu \nu}^{\phantom{\mu\nu}\!\kappa \lambda}(x) x_\lambda}{8\pi ^2x^4}j^c_\kappa (0) +\cdots ,}[JmuJnuOPE]
where $f^{ab}_c$ are the group structure constants, and $kd^{ab}_c$ is the coefficient of the $\Tr G^3$ 't Hooft anomaly.
The leading terms in the stress-tensor self-OPE are more involved to write out, because of all the indices, see \rcite{Osborn:1993cr}.
The terms $\sim 1/x^n$ for integer $n$ contribute to the conformal anomaly $\vev{T^{\phantom{\mu}\!\mu} _\mu}$ when the operators are coupled to background sources, see e.g.\ \rcite{Petkou:1999fv} for a nice discussion.  In particular, $\tau ^{ab}=\tau \delta ^{ab}$ gives the contribution to $\vev{T_\mu ^{\phantom{\mu}\!\mu}}$ when $j_\mu ^a(x)$ are coupled to external sources $A_\mu ^a(x)$, which shows that $\tau$ gives the contribution to the one-loop beta function for the gauge coupling if the $G$ symmetry is weakly gauged.

\newsec{4d \texorpdfstring{${\cal N}=1$}{N=1} SCFT primaries, descendants, and OPEs}[scOPE]

The ${\cal N}=1$ superconformal algebra (isomorphic to $SU(2,2|1)$) extends the conformal algebra with the supercharges $Q_\alpha$ and $\bar{Q}_{\dot \alpha}$, the superconformal supercharges, $S^\alpha$ and $\bar{S}^{\dot{\alpha}}$, and the $U(1)_R$-generator, $R$.  (See appendix \ref{SCA} for more details about the algebra.)

Representations are formed by regarding $P_\mu$, $Q_\alpha$, and $\bar{Q}_{\dot \alpha}$ as the raising operators, and $K_\mu$, $S^\alpha$, $\bar{S}^{\dot \alpha}$ as the corresponding lowering operators.  If an operator $\CO$ has $(\Delta, r)$ for its operator dimension and R-charge, respectively, then $Q_\alpha (\CO)\equiv [Q_\alpha, \CO\}$ has $(\Delta +\half, r-1)$ and e.g.\ $S^\alpha (\CO)\equiv [S^\alpha, \CO\}$ has $(\Delta -\half, r+1)$.
\begin{figure}[ht]
\centering
\includegraphics{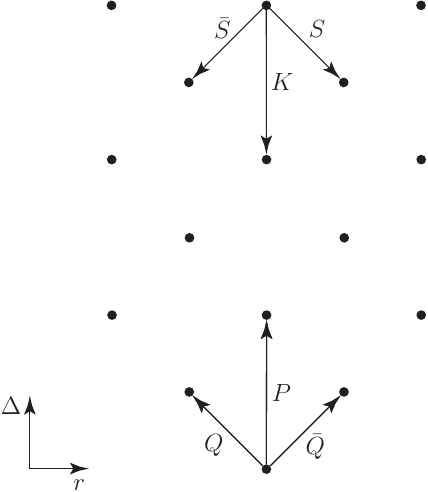}
\caption{A representation of the superconformal group.}\label{reps}
\end{figure}
The superconformal quasi-primary operators are at the bottom of the representations, annihilated by all lowering operators at the origin, $x^\mu =0$.   Each superconformal quasi-primary has a tower of superconformal descendant operators above it, obtained by acting with the raising operators; this is represented by the dots in Fig. \ref{reps}, with the superconformal quasi-primary operator at the bottom.\foot{In special cases some superconformal descendants are also primaries, i.e.\ annihilated by the lowering operators.  Such operators are zero-norm null states, that must be set to zero, leading to a truncated representation.  Examples are chiral primary operators $\CO$, where $\bar Q_{\dot \alpha}(\CO)\equiv[\bar Q_{\dot \alpha} , \CO]$ is null, and (semi-)conserved currents $J$, where $Q^2(J)\equiv \{Q^\alpha ,[Q_\alpha, J]\}$ is null.} The other operators on the bottom left and right edges,  e.g.\ $Q_\alpha (\CO)$,  are conformal primaries but superconformal descendants.

Every SCFT has a superconformal $U(1)_R$-plus-stress-energy-tensor supermultiplet\rcite{Ferrara:1974pz}
\eqn{
\CT_\mu(z)=j_\mu^R(x)+\theta ^\alpha {\cal S}_{\alpha \mu}(x)+\bar \theta ^{\dot \alpha}\bar {\cal S}_{\dot \alpha \mu}(x)
+2\theta\sigma^\nu\bar\theta T_{\nu\mu}(x)+\cdots,}[FZis]
where the $\cdots$ are derivative terms, determined by the conservation equation $\bar D^{\dot \alpha}\CT _{\alpha \dot \alpha}=0$. The primary component $j^R_\mu (x)=\CT _\mu |$ is the conserved superconformal $U(1)_R$ symmetry current, with  $\Delta _{j_\mu^R}=3$.
The supercurrents ${\cal S}_\mu ^\alpha (x)$, $\bar {\cal S}_\mu ^{\dot \alpha}(x)$, and the stress-energy tensor $T_{\mu \nu}(x)$ are among its descendants.   The leading short distance singular terms in the OPE of $\CT _\mu (z)$ with other operators, including itself, have coefficients with interesting universality \rcite{Anselmi:1996dd} interpretations, fixed in terms of the dimension and R-charges of the operators, 't Hooft anomalies, and the central charges $a$ and $c$.  The supersymmetry relations among the $j_R^\mu$ and $T_{\mu \nu}$ operators in \FZis\ then yields the relations of \rcite{Anselmi:1997am} and \rcite{Osborn:1998qu} between the central charges and the $U(1)_R$ 't Hooft anomalies.

Knowing how the superconformal generators act on the operator representations at $x^\mu =0$, their action at a general point $x^\mu$ follows from $\CO ^I(x)=e^{-iP\cdot x}\CO ^I(0)e^{iP\cdot x}$ and the algebra. For example, for a scalar superconformal primary, it follows that
\eqn{[S^\alpha, \CO(x)]=i x\cdot \bar \sigma^{\dot \alpha \alpha} [\bar Q_{\dot \alpha}, \CO(x)] .}[Sact]
As another example, raising and then lowering a scalar superconformal primary yields
\eqn{S^\beta Q_\alpha (\CO (x))=2(\sigma^{\mu\nu\phantom{\!\alpha}\!\beta}_{\phantom{\mu\nu}\!\alpha}x_{[\mu}\partial_{\nu]}+\delta_\alpha^{\phantom{\alpha}\!\beta}x\cdot\partial)\CO(x)-ix\cdot \bar \sigma^{\dot \alpha \beta}Q_\alpha \bar Q_{\dot \alpha} (\CO(x))+(2\Delta _\CO +3r_\CO)\delta_\alpha^{\phantom{\alpha}\!\beta}\CO(x), }[SQO]
where, again, we define $S^\beta Q_\alpha (\CO (x))\equiv \{S^\beta ,[Q_\alpha, \CO (x)]\}$.

Considering \SQO at $x^\mu =0$, it's seen that   $Q^\alpha(\CO (0))$ is null only if  $\Delta _\CO = -\frac{3}{2}r_\CO$; these are the anti-chiral primaries.  Similarly, it follows from $S^\alpha Q^2(\CO(0))=2[2(2-\Delta _\CO)-3r_\CO ]Q^\alpha(\CO (0))$, that $Q^2 (\CO)$ is null only if $\Delta _\CO= 2-\frac{3}{2}r_\CO$.
Likewise, $\bar Q^2(\CO)(0)$ is null only if
 $\Delta _\CO=2+\frac{3}{2}r_\CO$.  Conserved current operators satisfy both conditions,
 \eqn{ Q^2(J(x))=\bar{Q}^2 (J(x))=0,}[CurCons]
and so $\Delta _J=2$ and $r_J=0$.     The scalar primary operator $J(x)$ has the conserved current $j_\mu$ as a superpartner descendant, $j_\mu (x)=-\frac{1}{4}\bar \sigma _\mu ^{\dot \alpha \alpha}[Q_\alpha, \bar Q_{\dot \alpha}]J(x)$.

One might anticipate that, much as in \primOPE, the OPE for all operators is completely determined by those for the superconformal primaries,
\eqn{\CO _i (x)\CO _j(0) \stackrel{?}{=}\sum _{\substack{\rm{sprimary}\\ \CO_k}}\frac{c_{ij}^k}{(x^2)^{\frac{1}{2}(\Delta _i +\Delta _j-\Delta _k)}}F_{ij}^{k}(x,P, Q, \bar Q)\CO _k(0),}[sprimOPE]
where ``sprimary'' is shorthand for ``superconformal primary'', with the superconformal descendant OPE coefficients completely determined from those of the superconformal primaries.
But as we mentioned after \JJJsope, this is generally incorrect.  This is already known, but perhaps not widely so.  We can illustrate an example of from what we've discussed so far: consider the OPE $\CO^\dagger (x) \CO(x)$, where $\CO$ is a scalar operator with superconformal $U(1)_R$ charge $r_\CO$ and dimension $\Delta _\CO$.  It follows from \OOjT\ that
\eqn{\CO^\dagger (x)\CO(0)\supset -i r_\CO \frac{\pi ^2}{8c}\frac{c_{\CO ^\dagger \CO}x_\mu}{x^{2(\Delta _\CO -1)}}j_R^\mu (0)+\Delta _{\CO} \frac{\pi ^2}{60c} \frac{c_{\CO ^\dagger \CO}x^\mu x^\nu}{x^{2(\Delta _\CO -1)}}T_{\mu \nu}(0)+\cdots,}[OORT]
where we used the supersymmetry relation between the coefficient $\tau _{RR}$ of the $j_R^\mu$ two-point function and the conformal anomaly $c$, $\tau _{RR}=16c/3$ (see e.g.\ \rcite{Barnes:2005bm}).   Equivalently,
\eqn{{\cal T}_\mu (z) \CO (0)\supset \left(-i r_\CO \frac{x_\mu}{2\pi ^2 x^4}-4 \Delta _\CO\frac{1}{3\pi ^2 x^6}\theta \sigma ^\nu \bar \theta (x_\mu x_\nu-\tfrac14 x^2\eta_{\mu \nu})\right)\CO (0)+\cdots.}[FZOope]

For a general operator $\CO$, the coefficients $r_\CO$ and $\Delta _\CO$ in \OORT\ or \FZOope\ are not proportional to each other (only for chiral or anti-chiral primaries is there a fixed proportionality).
So, for general operators $\CO$, the two terms on the RHS of  \OORT\ have two independent OPE coefficients, for the primary operator, $j_R^\mu$, and its super-descendant, $T^{\mu \nu}$.   This illustrates that \sprimOPE\ can not hold with any universal functions $F_{ij}^k$.  Generally,  the coefficients  of the $Q$ and $\bar Q$ descendant terms in $F$ in \sprimOPE\ are independent coefficients, not fixed by the symmetries.  This all follows from the
 general superpace analysis of Osborn \rcite{Osborn:1998qu}, that we'll now review.


\subsec{Two and three-point functions using the superspace analysis of \texorpdfstring{\rcite{Osborn:1998qu}}{Osborn}}[Osb]

Operators are labeled by $(j, \bar \jmath, q, \bar q)$, where $(j, \bar \jmath)$ are the Lorentz spins, $q\equiv \half (\Delta +\frac{3}{2} r)$ and ${\bar q}\equiv \half (\Delta -\frac{3}{2}r)$, where $\Delta$ is the operator's dimension and $r$ its R-charge.  Chiral operators have $\bar q=0$, real operators have $q=\bar q=\half \Delta$, and conserved currents have $q=\bar q=1$.   The form of two-point functions of arbitrary superconformal primaries  is completely fixed in  \rcite{Osborn:1998qu} by superconformal invariance, up to overall coefficients $c_{k\bar k}$ (which could be set to $\delta _{k\bar k}$ by choice of operator normalization for some operators (but not $J$ or ${\cal T}_\mu$)):
\eqn{\vev{\CO ^{i_3}_k(z_2)\bar{\CO}_{\bar k}^{\bar \imath_3}(z_3)}=c_{k\bar k}\frac{I^{i_3\bar \imath_3}(x_{2\bar 3}, x_{\bar 2 3})}{\osbx{\bar 23}{2\bar q_3}\osbx{\bar 32}{2q_3}}.}[OsbornOO]
Here $z_i$ denotes superspace coordinates, $z_i=(x_i^\mu, \theta _i ^\alpha, \bar \theta _i^{\dot \alpha})$, $x_{ij}^\mu=x_i^\mu-x_j^\mu$, $\theta_{ij}^\alpha=\theta_i^\alpha-\theta_j^\alpha$, and
\eqn{
\osbx{\bar{\imath}j}{\,\mu}=x_{ij}^\mu-i\theta_i\sigma^\mu\bar{\theta}_j+i\theta_j\sigma^\mu\bar{\theta}_{i}-i\theta_{ij}\sigma^\mu\bar{\theta}_{ij}.
}
$I^{i_3\bar \imath_3}(x_{2\bar 3}, x_{\bar 2 3})$, where $\osbx{i \bar{\jmath}}{\,\mu}=-\osbx{\bar{\jmath}i}{\,\mu}$, is a bilocal invariant tensor in the spin indices $i_3$, $\bar \imath_3$, reducing to 1 for scalars (see \rcite{Osborn:1998qu} for the explicit expression).

The form of three-point functions is determined in \rcite{Osborn:1998qu} to be
\eqn{\vev{\CO _1^{i_1}(z_1)\CO _2^{i_2}(z_2) \CO_3^{\dagger i_3}(z_3)}=\frac{I_1^{i_1 \bar \imath_1} (x_{1\bar 3}, x_{\bar 13})I_2^{i_2\bar \imath_2}(x_{2\bar 3},x_{\bar 23})}{\osbx{\bar 13}{2\bar q_1}\osbx{\bar 31}{2q_1}\osbx{\bar 23}{2\bar q_2}\osbx{\bar 32}{2q_2}}t_{\bar \imath_1 \bar \imath_2}^{i_3}(X_3, \Theta _3, \bar \Theta _3).}[OsbornOOO]
We called the third operator $\CO ^\dagger _3$ because we're eventually interested in the OPE, $\CO _1\CO_2 \sim \CO_3$.    $X_3^\mu$ is a 4-vector formed from the superspace coordinates $z_{i=1,2,3}=(x_i,\theta_i,\bar{\theta}_i)$ \rcite{Osborn:1998qu},
\eqn{{\rm{X}}_3\equiv \frac{{\rm x}_{2\bar 1}\tilde {\rm x}_{\bar 1 2}{\rm x}_{2\bar 3}}{\osbx{\bar 13}{2}\osbx{\bar 32}{2}}; \qquad ({\rm X}_3)_{\alpha \dot \alpha}=\sigma _{\mu\alpha \dot \alpha} X_3^\mu, \qquad \tilde{\rm x}^{\dot{\alpha}\alpha}=\epsilon^{\alpha\beta}\epsilon^{\dot{\alpha}\dot{\beta}}{\rm x}_{\beta\dot{\beta}}.}[Xdefn]
The spinor quantities in \OsbornOOO\ are given by
\eqn{\Theta _3\equiv i\left(\frac{1}{\osbx{\bar 13}{2}}{\rm x}_{3\bar 1}\bar \theta _{31}-\frac{1}{\osbx{\bar 2 3}{2}}{\rm x}_{3\bar 2}\bar \theta _{32}\right),\qquad \bar \Theta _3\equiv i\left(\frac{1}{\osbx{\bar 31}{2}} \theta _{31}{\rm x}_{1\bar 3}-\frac{1}{\osbx{\bar 32}{2}}\bar \theta _{32}{\rm x}_{2\bar 3}\right), }[Thetais]
which are nilpotent, they vanish upon setting the Grassmann coordinates to zero, and they don't have a direct analog in ordinary conformal theories.
$X_3^\mu$ is a superspace extension of the vector  $Z^\mu$ defined in \Zis, $Z^\mu = \frac{x_{31}^\mu}{r_{13}}-\frac{x_{32}^\mu}{r_{23}}$.
For example, setting the Grassmann part of the $z_{i=1,2}$ coordinates to zero, and defining $Y^{\mu \nu}\equiv \epsilon ^{\mu \nu}_{\phantom{\mu \nu}\!\rho \lambda} \frac{x_{13}^\rho x_{23}^\lambda}{r_{13} r_{23}}$, we find
\eqn{X_3^\mu |_{\theta _{i=1,2}=\bar \theta _{i=1,2}=0}=Z^\mu +[i(Z^2\eta ^{\mu \nu}-2Z^\mu Z^\nu) +{\color{red}2Y^{\mu\nu}}]\theta _3\sigma_\nu \bar \theta _3+Z^2\left({\color{red}\frac{x_{12}^\mu}{r_{12}}}-Z^\mu\right)\theta _3^2\bar\theta _3^2}[Xexp]
(the red terms will drop out).  The function $t$ in \OsbornOOO\ is generally under-determined, constrained only by a homogeneity condition corresponding to the scale and R-charges:
\eqn{t_{\bar \imath_1\bar \imath_2}^{i_3}(\lambda \bar \lambda X, \lambda \Theta , \bar \lambda \Theta)=\lambda ^{2a}\bar \lambda ^{2\bar a}t_{\bar \imath_1\bar \imath_2}^{i_3}(X, \Theta , \bar \Theta),}[tscaling]
with
\eqn{a-2\bar a=\bar q_1+\bar q_2-q_3, \qquad \bar a-2a=q_1+q_2-\bar q_3.}[ais]

Conformal three-point functions of primaries have a fully-determined dependence on the operator locations, which can be viewed as a consequence of the fact that ordinary
conformal symmetry transformations can be used to map any three points to any three other points.  But superconformal symmetry does not suffice to map three super-positions $z_i$ to wherever one pleases, and that is related to the existence of the $\Theta$, $\bar \Theta$ in \OsbornOOO\ and \Thetais.  Indeed,  supertranslations can be used to set, say, $z_1=0$ and superconformal transformations can be used to map, say, $x_2=\infty$ and $\theta _2=\bar \theta _2=0$.  Then we are left with the $z_3\equiv z$ superspace coordinate, which we can act on with ordinary rotations, $U(1)_R$ rotations, and scale transformations.  With these mappings, $X_3^\mu$ is given by \Xexp\ with $Z^\mu \to x^\mu /x^2$, $x_{12}^\mu /r_{12}\to 0$ and $Y^{\mu \nu}\to 0$.  The nilpotent quantities \Thetais\ map to
\eqn{\Theta \to \frac{i}{x^2}({\rm x}-i\theta\bar{\theta})\bar{\theta}, \qquad \bar \Theta \to \frac{-i}{x^2}\theta({\rm x}+i\theta\bar{\theta}).}[Thetamap]

The existence of $\Theta _3$ and $\bar \Theta _3$, and the fact that $t$'s dependence on them is generally under-determined by \tscaling,  implies that the three-point functions of superconformal primaries are generally \emph{insufficient} to fully determine those of their superconformal descendants.  The superconformal primary three-point functions are extracted by setting the Grassmann coordinates to zero, but that's generally insufficient to determine the $\Theta _3$ and $\bar \Theta _3$ dependence (since they then vanish), which is  needed to determine the three-point function of general superconformal descendants.    So the OPE coefficients of superconformal primaries generally do not fully determine those of their superconformal descendants.

This general ambiguity in the function $t(X, \Theta, \bar \Theta)$ is eliminated only in special cases, when some of the three operators are in reduced superconformal representations, with null states, e.g.\ chiral primaries, anti-chiral primaries, or conserved currents.  Superspace derivatives on the operators $\CO _i$ in \OsbornOOO\ can be converted into differential operators acting on the function $t(X_3, \Theta _3, \bar\Theta _3)$, and so constraints on the operators lead to corresponding constraints on the function $t(X_3, \Theta _3, \bar\Theta _3)$.   In particular, acting on say $\CO _1$, one replaces $D_\alpha \to {\cal D}_\alpha$ and ${\bar{D}}_{\dot{\alpha}}\to {\bar{{\cal D}}}_{\dot{\alpha}}$, which act on $t(X, \Theta, \bar \Theta)$ as \rcite{Osborn:1998qu}
\eqn{{\cal D}_\alpha = \frac{\partial}{\partial \Theta ^\alpha }-2i(\sigma ^\mu\bar \Theta )_\alpha \frac{\partial}{\partial X^\mu}, \qquad \bar {\cal D}_{\dot \alpha}=-\frac{\partial}{\partial \bar \Theta^{\dot \alpha}},}[dtops]
with $\bar X=X-4i\Theta \bar \Theta$.  As examples, we'll first review the cases that have been discussed in the literature, where $\CO_1$ and $\CO_2$ are chiral or anti-chiral operators.  In the following section, we'll consider  our case of interest: conserved currents.

\subsec{Review of chiral-chiral OPEs \texorpdfstring{\rcite{Dolan:2000uw,Poland:2010wg,Vichi:2011ux}}{Dolan and Poland}}

Take the operators $\CO _1$ and $\CO _2$ in the three-point function \OsbornOOO\ to both be chiral primaries, which we'll write as $\CO _i=\phi _i$.   The condition $\bar D_{\dot{\alpha}} \phi _1=0$ implies that $\bar {\cal D}_{\dot{\alpha}} t=0$ for the operator in \dtops, with a similar condition for $\bar D_{\dot{\alpha}} \phi _2$.  If we take $\phi_1$ and $\phi _2$ to be the same operator, the latter condition is accounted for by the  $z_1\leftrightarrow z_2$ symmetry, which implies
\eqn{t(X_3, \Theta _3, \bar \Theta _3)=t(-\bar X_3, -\Theta _3, -\bar \Theta _3).}[tsymm]
The solutions for $t(X _3, \Theta _3, \bar \Theta _3)$ are \rcite{Dolan:2000uw,Poland:2010wg,Vichi:2011ux}
\eqna{
t&\sim \text{constant},\\
t&\sim \bar \Theta _3 \bar X_3^{\Delta _\CO -\Delta _i-\Delta _j-1-\ell}\bar X_3^{\mu _1}\cdots \bar X_3^{\mu _\ell},\\
t&\sim \bar \Theta _3^2  \bar X_3^{\Delta_{\CO _3}-\Delta _i-\Delta _j-\ell+1}X_3^{\mu_1}\cdots X_3^{\mu_\ell}.}
The case $t\sim $ constant implies that the operator $\CO _3$ in the three-point function \OsbornOOO\ is also chiral, $\CO _3=\phi _k$, with $R(\CO _3^I)=R(\phi _i)+R(\phi _j)-2$; this is the chiral ring.  The other two cases for $t$ have factors of $\bar \Theta _3$ and $\bar \Theta _3^2$, corresponding to operators $\CO _3$ in \OsbornOOO\ that are $\bar Q_\alpha$ and $\bar Q^2$ exact (hence trivial in the chiral ring, but nevertheless important for non-holomorphic considerations). Correspondingly, the possible terms in the OPE are
\eqn{\phi _i(x)\phi _j(0)=c_{ij}^kC(x,P)\phi _k(0)+\sum _{\CO^I} c_{ij}^{\CO^I}\bar Q C_I(x, P)\CO ^I(0)+\sum _{\CO^J} c_{ij}^{\CO^J}\bar Q^2 C_J(x, P)\CO ^J(0),}[chiralring]
where $c_{ij}^k$, $c_{ij}^{\CO_I}$, and $c_{ij}^{\CO _J}$ are constant OPE coefficients.  The operators $\CO _J$ in \chiralring\ have even spin, $\ell =2j_1=2j_2$, and $R(\CO _J)=\frac{2}{3}(\Delta _i+\Delta _j)-2$ (so unitarity requires $\Delta _\CO\geq |\frac{3}{2}R_\CO|+\ell +2$).  To give a simple example, consider a theory with a chiral superfield $\Phi$, $K=\bar \Phi \Phi$ and $W=\lambda \Phi ^{n+1}/(n+1)$. Then, the equation of motion $\Phi ^n=-\bar Q ^2\bar \Phi /\lambda$ illustrates the last term in \chiralring.  The $\CO _I$ possibility in \chiralring runs only over superconformal primaries with $R(\CO _J)=\frac{2}{3}(\Delta _i+\Delta _j)-1$, spins $(j_1, j_2)= (\half (\ell +1), \half \ell)$, with $\ell $ odd for \tsymm, and $\Delta (\CO _J)=\Delta _i+\Delta _j+\ell +\half$, where $\Delta$ is fixed (saturating a unitarity bound) because the operator $\CO _J$ must be in a shortened multiplet to have both sides of \chiralring\ annihilated by $\bar Q_{\dot \alpha}$.

In \chiralring we have written just the first components of the superfields on the LHS.  The full superfield expression for the first term in \chiralring\ was worked out in \rcite{Dolan:2000uw}:
\eqn{\Phi _i(z_{1+})\Phi _j(z_{2+})\supset c_{ij}^k {\cal C}^{q_i, q_j}(z_{12+}, \partial _{z_2+}) \Phi _k (z_{2+}),}[chiralrings]
which has no $x_{12}$ singularity since $q_k=q_i+q_j$ for the chiral ring, and
\begin{multline*}{\cal C}^{q _1, q _2}(z_{12+}, \partial _{z_2+})\frac{1}{(x_{2+}-2i\theta _2\sigma \bar \theta -x_-)^{2q_1+2q_2}}=\\
\frac{1}{(x_{1+}-2i\theta _1\sigma \bar\theta-x_-)^{2q_1}(x_{2+}-2i\theta _2\sigma \bar \theta-x_-)^{2q_2}},
\end{multline*}
which was solved for in \rcite{Dolan:2000uw}\ in a superspace expansion in $\theta _{12}$, with components given by the functions $C^{ab}(x_{12}, \partial _{x_2})$ in \Ceqn.

\subsec{Review of chiral-anti-chiral OPE \texorpdfstring{\rcite{Poland:2010wg}}{Poland}}

Let the operators $\CO _1$ and $\CO_2$ in \OsbornOOO\ be chiral and anti-chiral respectively.  As in \rcite{Poland:2010wg}, for simplicity we'll take $\CO _1=\Phi$ and $\CO _2=\bar \Phi$, the conjugate field.  The conditions $\bar{\cal D}_{1,\dot \alpha}t=0$ and ${\cal D}_{2, \alpha}t=0$ then imply that the operator $\CO _3$  must be real and of integer spin $\ell =2j=2\bar \jmath$, with \rcite{Poland:2010wg}
\eqn{\vev{\Phi (z_{1+})\bar \Phi (z_{2-})\CO ^{\mu_1\dots \mu_\ell}(z_3)}\propto \frac{1}{\osbx{\bar 3 1}{2\Delta _\Phi}\osbx{\bar 2 3}{2\Delta _\Phi}}\bar X_3^{\Delta _{\CO}-2\Delta _\Phi -\ell}\bar X_3^{\mu_1}\cdots \bar X_3^{\mu_\ell}-\rm{traces}.}[ccbar]
The result \ccbar\ encodes interesting relations among the component OPE coefficients.   We will review this in some detail, following  \rcite{Poland:2010wg}, since many details will prove applicable for our case of interest, to be discussed in the next section.

Real operators $\CO ^{\mu_1\dots \mu_\ell}$ in \ccbar\ have a superspace expansion
\eqn{\CO ^{\mu _1\dots \mu _\ell}(x, \theta, \bar\theta)=A^{\mu _1\dots \mu _\ell}(x)+\xi _\mu B^{\mu \mu _1\dots \mu _\ell}(x)+\xi ^2 D^{\mu _1\dots \mu _\ell}(x)+\cdots,}[realo]
where $\xi _\mu \equiv \theta \sigma _\mu \bar \theta$ and $\cdots$ are operators with non-zero R-charge.  The $A$ component is primary, and the others others are all $A$'s descendants: defining $\Xi ^\mu\equiv \bar \sigma ^{\mu \dot \alpha \alpha}[Q_\alpha, \bar Q_{\dot \alpha}]$,
\eqn{B^{\mu \mu _1\dots \mu _\ell}=-\frac{1}{4}\Xi ^\mu A^{\mu _1\dots \mu _\ell}, \qquad D^{\mu _1\dots \mu _\ell}=-\frac{1}{64}\Xi _\mu B^{\mu \mu _1\dots \mu _\ell}-\frac{1}{16}\partial ^2 A^{\mu _1\dots \mu _\ell}.}[realox]
The operators $A^{\mu _1\dots \mu _\ell}$ and $D^{\mu _1\dots \mu _\ell}$ are irreducible spin-$\ell$ representations, while $B^{\mu \mu _1\dots \mu _\ell}$ decompose into $B^{\mu \mu _1\dots \mu _\ell}=M^{\mu \mu _1\dots \mu _\ell}+\frac{\ell ^2}{(\ell +1)^2}\eta ^{\mu \mu _1}N^{\mu _2\dots \mu _\ell}+L^{\mu \mu _1\dots \mu _\ell}$, where $M$ (called $J$ in \rcite{Poland:2010wg}) is a spin $\ell +1$ operator, $N$ is a spin $\ell -1$ operator, and $L=L_++L_-$, with $L_\pm$ in the $(\half \ell \pm \half, \half \ell \mp \half)$ representation of $SU(2)\times SU(2)$.   The operators $B$ and $D$ can be decomposed into conformal primary and descendant contributions \rcite{Poland:2010wg}, with $M^{\mu \mu _1\dots \mu _\ell}_{\prim}=M^{\mu \mu _1\dots \mu _\ell}$, $N^{\mu _2\dots \mu _\ell}_{\prim}=N^{\mu _2\dots \mu _\ell}$, and ($P_\mu ^{\rm here}=-iP_\mu ^{\rm there}$, as we prefer Hermitan generators)
\eqna{L^{\mu \mu _1\dots \mu _\ell}_\prim&=L^{\mu \mu _1\dots \mu _\ell}-\frac{\ell} {4(\Delta -1)}\epsilon ^{\mu \mu _1}_{\phantom{\mu \mu _1}\! \nu\rho}iP^\nu  A^{\rho\mu _2\dots \mu _\ell},\cr
D^{\mu _1\dots \mu _\ell}_\prim &= D^{\mu _1\dots \mu _\ell}-\frac{\ell (\ell +1)-(\Delta -1)}{8(\Delta -1)^2}P^2 A^{\mu _1\dots \mu _\ell}+\frac{\ell ^2}{4(\Delta -1)^2}P_\nu P^{\mu _1}A^{\nu \mu _2\dots \mu _\ell}\cr &\quad -\frac{\ell}{4(\Delta -1)}\epsilon ^{\mu _1\nu}_{\phantom{\mu_1\nu}\!\rho \sigma}iP_\nu L^{\rho \sigma \mu _2\dots \mu _\ell}.}[LDprim]

Setting for example $\theta _1=\theta _2=\bar \theta _1=\bar \theta _2=0$ in \ccbar\ to extract the three-point functions for $\phi =\Phi |$ and $\bar \phi =\bar \Phi |$, it is found that \rcite{Poland:2010wg}
\eqna{\vev{\phi \phi ^* A^{\mu _1\dots \mu _\ell}}&=c_{\phi \phi ^* \CO_\ell} \frac{Z^{\Delta  -\ell}}{r_{12}^{\Delta _\Phi}}Z^{\mu _1}\cdots Z^{\mu _\ell},\cr
\vev{\phi \phi ^* M _\prim ^{\mu \mu _1\dots \mu _\ell}}&=ic_{\phi \phi ^* \CO _\ell} (\Delta +\ell) \frac{Z^{\Delta  -\ell}}{r_{12}^{\Delta _\Phi}}Z^{\mu} Z^{\mu _1}\cdots Z^{\mu _\ell},\cr
\vev{\phi \phi ^* N^{\mu _2\dots \mu _\ell}_\prim}&=ic_{\phi \phi ^* \CO _\ell} \frac{(\ell +1)(\Delta -\ell -2)}{2\ell}\frac{Z^{\Delta +2 -\ell}}{r_{12}^{\Delta _\Phi}}Z^{\mu _2}\cdots Z^{\mu _\ell},\cr
\vev{\phi \phi ^* L^{\mu \mu _1\cdots \mu _\ell}_\prim} &=0,\cr
\vev{\phi \phi ^* D_\prim^{a_1\cdots a_\ell}}&=-c_{\phi \phi ^*\CO _\ell} \frac{\Delta(\Delta+\ell)(\Delta -\ell -2)}{8(\Delta -1)} \frac{Z^{\Delta +2  -\ell}}{r_{12}^{\Delta _\Phi}}Z^{\mu _1}\cdots Z^{\mu _\ell},}[ADrelni]
where $Z$ is the quantity in \Zis\ and the products like $Z^{\mu _1}\cdots Z^{\mu _\ell}$ are to be understood as symmetrized traceless.   The primary three-point functions \ADrelni\ indeed have the form \OOOspin, involving only the coordinate $Z^\mu$.  Indeed,  $\vev{\phi \phi ^* L_{{\rm prim}}}$ had to vanish, since it's impossible to form something with $L$'s Lorentz structure using only $Z^\mu$.  The upshot of \ADrelni\ is that the  coefficient $c_{\phi \phi ^*\CO _\ell}=c_{\phi \phi ^*A_\ell}$ of the superconformal primary $A$ indeed completely determines those of the descendants, $M$, $N$, $L_\text{\prim}$, and $D_\text{prim}$:
\eqna{c_{\phi \phi ^*{M_{\ell+1}}}&=i(\Delta +\ell)c_{\phi \phi ^*A_{\ell}}, \cr
c_{\phi \phi ^*N_{\ell-1}}&=i\frac{(\ell +1)(\Delta -\ell -2)}{2\ell}c_{\phi \phi ^*A _{\ell}},\cr
\qquad \qquad c_{\phi \phi ^*L_\prim}&=0,\cr
c_{\phi \phi ^*D_{\ell;\text{prim}}}&=-\frac{\Delta (\Delta +\ell)(\Delta -\ell -2)}{8(\Delta -1)}c_{\phi \phi ^*A_\ell}.}[PSDrelns]

To convert \PSDrelns\ to relations among  the OPE coefficients, we can use $c_{ij}^k=c_{ijk'}g^{k'k}$ \craise, and the relations among the two-point function normalizations.
 The
two-point function of $A_\ell$ is proportional to
\eqn{\langle A^{\nu _1\dots \nu _\ell}|A^{\mu _1\dots \mu _\ell}\rangle \sim \text{symmetrize}(\eta ^{\mu _1\nu_1}\cdots \eta ^{\mu _\ell \nu_\ell})-\text{traces}\equiv {\cal I}_\ell ^{\mu _1\dots \mu _\ell; \nu _1\dots \nu _\ell}.}
Likewise, the two-point functions of $M_{\ell+1}$, $N_{\ell-1}$, and $D_{\ell}$ are proportional to ${\cal I}_{\ell +1}$, ${\cal I}_{\ell -1}$, and ${\cal I}_\ell$, respectively, and $\langle L_\prim ^{\nu \nu _1\dots \nu _\ell}|L_\prim ^{\mu \mu _1\dots \mu _\ell}\rangle\sim \eta ^{\mu \nu} {\cal I}_\ell ^{\mu _1\dots \mu _\ell; \nu _1\dots \nu _\ell}$.   The proportionality factors for the two-point function normalization of the super-descendants, relative to the primary component, are given by  \rcite{Poland:2010wg}
\eqna{c_{M_{\ell+1} M_{\ell +1}} &=2(\Delta +\ell)(\Delta +\ell +1)c_{A_\ell A_\ell},\cr
c_{N_{\ell -1}N_{\ell-1}} &=\frac{2(\ell +1)^2(\Delta -\ell-2)(\Delta -\ell -1)}{\ell ^2}c_{A_\ell A_\ell},\cr
c_{L_\prim L_\prim}&=\frac{8\ell ^2 \Delta (\Delta +\ell)(\Delta -\ell -2)}{(\ell +1)^2(\Delta -1)}c_{A_\ell A_\ell},  \cr
c_{D_{\ell; \prim}D_{\ell; \prim}}&=\frac{\Delta ^2(\Delta -\ell -2)(\Delta -\ell -1)(\Delta +\ell)(\Delta +\ell +1)}{4(\Delta -1)^2}c_{A_\ell A_\ell}, }[ADnorms]
where the factor $c_{A_\ell A_\ell}$ could be set to one by choice of normalization of $\CO _\ell$. Note that when the unitarity bound $\Delta \geq \ell +2$ is saturated, the norm \ADnorms\ of $N_{\ell-1}$, $L_{\prim}$, and $D_{\ell; \prim}$ all vanish; indeed, these components of the supermultiplet vanish when the unitarity bound is saturated---the supermultiplet is shortened.

\subsec{Another example: the \texorpdfstring{$\vev{\CO \CO ^\dagger \CT _\mu}$}{<O Odagger Tmu>} three-point function}

As another example of applying the general formalism of \rcite{Osborn:1998qu}, we can consider the three-point function the stress-energy tensor supermultiplet $\CT _\mu$ \FZis\ with a scalar superfield $\CO$ and its conjugate $\CO ^\dagger$.  For the case $\CO =\Phi$ a chiral operator, $q=\Delta _\Phi ={\frac32}r_\Phi$, $\bar q=0$, the result was given in \rcite{Osborn:1998qu},
\eqn{\vev{\Phi (z_{1+})\bar \Phi (z_{2-})\CT ^\mu (z_3)}=-ir_\Phi\frac{c_{\phi \bar \phi}}{2\pi ^2}\frac{1}{\osbx{\bar 3 1}{2q}\osbx{\bar 23}{2q}}\frac{\bar X_3^\mu }{\bar X_3^{2(q-1)}},}[cat]
where $c_{\phi \bar \phi}$ is the $\vev{\phi \bar \phi}$  two-point function normalization, and the coefficient in \cat\ is fixed by the condition that the OPE reproduces the correct $U(1)_R$ charge, as in \OOJi.  This is a special case of \ccbar, where we take $\Delta _\CO =3$ and $\ell =1$ to get $\CO ^\mu \to \CT ^\mu$.   So $c_{\phi \phi ^*j_R^\mu}=-i r_\Phi c_{\phi \phi ^*}/2\pi ^2$ and then \PSDrelns, with $M_{\mu \nu}=2T_{\mu \nu}$ (see \FZis) gives $c_{\phi \phi ^* T_{\mu \nu}}=r_\Phi c_{\phi \phi ^*}/\pi ^2$, which fits with \TOope\ and $\Delta = \frac{3}{2}|r_\Phi|$ for chiral and anti-chiral operators.

As another example, we consider the case where the operator $\CO$ is real, $\CO = \CO ^\dagger$, so $q_\CO =\bar q _\CO=\half \Delta _\CO$, and $R_\CO =0$.  Using  \OsbornOOO,  \tscaling, and the $z_1\leftrightarrow z_2$ symmetry we find
\eqn{\vev{\CO (z_1)\CO(z_2)\CT ^\mu (z_3)}=\frac{-\Delta _\CO c_{\CO \CO}}{6\pi ^2(\osbx{\bar 13}{2}\osbx{\bar 31}{2}\osbx{\bar 23}{2}\osbx{\bar 32}{2})^{\half \Delta _\CO }(X _3\cdot \bar X_3)^{\Delta_{\CO} -1 }}\left[ X_-^\mu +2 \frac{(X_-\cdot X_+)X_+^\mu }{X_3\cdot \bar X _3} \right],}[rrt]
where $X_+^\mu \equiv \half (X_3^\mu +\bar X_3^\mu)$ is a vector that's odd under the $z_1\leftrightarrow z_2$ operation in \tsymm, and $X_-^\mu \equiv i(X_3^\mu -\bar X _3^\mu)\equiv -4\Theta _3\sigma^\mu\bar \Theta _3$ is a (nilpotent) vector that's even under the $\mathbb{Z}_2$.    So $X_3\cdot \bar X_3=X_+^2+4\Theta_3 ^2\bar \Theta_3 ^2$.   The relative factor of two between the two terms in the sum on the RHS is determined by the condition $D_\alpha T^{\alpha \dot \alpha}=\bar D_{\dot \alpha }T^{\alpha \dot \alpha}=0$, and the overall normalization by \TOope. As a special case of \rrt, the three-point function of two conserved currents and the stress tensor is
\eqn{ \vev{\CJ (z_1) \CJ (z_2) \CT ^\mu (z_3)}=-\frac{\tau _{JJ}}{48\pi ^6 \osbx{\bar 13}{2}\osbx{\bar 31}{2}\osbx{\bar 23}{2}\osbx{\bar 32}{2}X _3\cdot \bar X_3}  \left[ X_-^\mu +2 \frac{(X_-\cdot X_+)X_+^\mu }{X_3\cdot \bar X _3} \right],}[JJt]
Comparing the $\vev{J(x_1) J(x_2) T^{\mu \nu}(x_3)}$ and the $\vev{j^\rho  (x_1) j^\sigma  (x_2) j_R^\mu(x_3)}$ components encoded in \JJt\ leads to the relation  $\tau _{JJ}=-3\Tr F^2R$, giving the current two-point function coefficient $\tau$ in \jjex\ as a 't Hooft anomaly.

\newsec{Our case of interest: the current-current OPE}[JJOPEsec]

We now consider the OPE of two $\Delta =2$ conserved-current primary operators,
\eqn{ J(x)J(0) =\sum _{\ell =0}^\infty \sum _{\substack{\rm{sprimary}\\ \CO^{(\ell)}_k}}\frac{c_{JJ}^k}{(x^2)^{\frac{1}{2}(4-\Delta _k)}}F_{JJ}^{k}(x,P, Q, \bar Q)^{(\ell)}\CO ^{(\ell)}_k(0),}[JJsOPE]
where $\CO _k^{(\ell)}$ are superconformal primaries, of dimension $\Delta _k$ and spin $\ell$, and we will show that the $\CO _k^{(\ell)}$ are necessarily real, of $U(1)_R$-charge zero.
For simplicity, we consider $U(1)$ currents.  The LHS of \JJsOPE\ is then symmetric under
exchanging the operators, and hence $x^\mu \to -x^\mu$, so only even spin operators can contribute on the RHS of the OPE.   For non-Abelian groups, odd spin components can appear on the RHS of $J_a(x) J_b(0)$, with coefficients proportional to $f_{abc}$ as in \jjex.   We discuss how to determine the $F_{JJ}^k(x,P, Q, \bar Q)^{(\ell)}$ from the condition of superconformal covariance, combined with $J$'s current conservation.

The OPE result \JJsOPE\  for the bottom component of the supercurrent multiplet will determine the OPE coefficients of its superconformal descendants, in  particular of \eqn{
j_\alpha(x) =Q_\alpha(J(x)),\qquad
j_\mu(x) =-\tfrac{1}{4}\Xi _\mu (J(x)),
}[Jdesc]
where $\Xi _\mu \equiv \bar{\sigma}_\mu^{\dot{\alpha}\alpha}[Q_\alpha ,\bar{Q}_{\dot{\alpha}}]$.
We can use $Q_\alpha$ and $\bar Q_{\dot \alpha}$ to map from the primary $J$, to its descendants, as in \Jdesc.  We can also map in the opposite direction, by using the $S^\alpha$ and $\bar S^{\dot \alpha}$ superconformal supercharges, which act on the primary component as
\eqn{S^\alpha(J(x))=ix\cdot \bar \sigma^{\dot \alpha \alpha}\bar Q_{\dot \alpha}(J(x)), \qquad \bar S^{\dot \alpha}(J(x))=-ix\cdot \bar\sigma^{\dot \alpha \alpha}Q_\alpha(J(x)),}[SJ]
vanishing at the origin.  Acting on the descendants as in \SQO with  $\Delta _J=2$ and $r_J=0$, we find
\twoseqn{S^\alpha (j_\alpha(x))&=-ix\cdot \bar \sigma^{\dot \alpha \alpha}Q_\alpha \bar Q_{\dot \alpha}(J(x))+4(x\cdot\partial+2)J(x),}[Sj]{S^\alpha(j^\mu(x))&=3\bar{\sigma}^{\mu\dot{\alpha}\alpha}\bar{\jmath}_{\dot{\alpha}}(x)-2x\cdot\bar{\sigma}^{\dot{\alpha}\alpha}\bar{\sigma}^{\mu\nu\dot{\beta}}_{\phantom{\mu\nu\dot{\beta}}\!\dot{\alpha}}\partial_\nu\bar{\jmath}_{\dot{\beta}}(x).}[Sjmu]

\subsec{Using the algebra to find relations in the \texorpdfstring{$J(x)J(0)$}{J(x)J(0)} OPE}

In this subsection, we discuss how superconformal symmetry leads to relations for $J(x)J(0)$ by directly using the algebra.   The relations obtained this way alternatively follow from using the superspace formalism of \rcite{Osborn:1998qu}, which we will use in the next subsection.

When the superconformal generators act on the product $J(x)J(0)$, the product rule gives two terms, e.g.\  $Q_\alpha (J(x)J(0))=Q_\alpha (J(x))J(0)+J(x)Q_\alpha (J(0))$.  But for the
lowering operators, $S^\alpha$, $\bar S^{\dot \alpha}$ and $K_\mu$, the term where they act on
the primary  $J(0)$ vanishes, so e.g.\
\eqn{S^\alpha (J(x)J(0))=S^\alpha (J(x))J(0)=-ix\cdot \bar \sigma^{\dot \alpha \alpha}\bar \jmath_{\dot \alpha} (x) J(0).}[SJJ]
The $j_\alpha (x) J(0)$ OPE thus follows from the $J(x)J(0)$ OPE, with only superdescendants in $J(x)J(0)$ contributing to the OPE around the origin, since  superconformal primary terms are annihilated by $S^\alpha$ in \SJJ.

The relation \SJJ\ illustrates how the OPE $J(x)J(0)$ of the primary operators in the multiplet determine the OPEs of the descendants.  Additional relations follow because we are here considering conserved currents rather than generic operators, so $Q^2(J(x))=\bar Q^2(J(x))=0$.
For example, consider the $j^\alpha (x) j_\alpha (0)$ operator product, relevant for determining gaugino masses in general gauge mediation, which can be related to $J(x)J(0)$ as in \rcite{Buican:2008ws} (see appendix \ref{SCA} for a discussion about the sign)
\eqn{j^\alpha (x)j_\alpha (0)=\tfrac{1}{2} Q^2(J(x)J(0)).}[QQJJ]
In superconformal theories, this descendant operator product can also be related to the primary $J(x)J(0)$ by using \SJ\ as
\eqn{j_\alpha (x) j_\beta (0)=\frac{1}{x^2}Q_\beta (ix\cdot \sigma \bar S)_\alpha (J(x)J(0)).}[QSJ]
Again, $\bar S^{\dot \alpha}$ only acts on $J(x)$, and then  $Q_\beta$ only acts on $J(0)$ (since $Q^2(J(x))=0$).  Another interesting relation that follows from \SJ, combined with $Q^2(J(x))=\bar Q^2(J(x))=0$, is
 \eqn{S^\alpha S^\beta (J(x)J(0))=\bar S^{\dot \alpha}\bar S^{\dot \beta}(J(x)J(0))=0.}[SSJ]
The relations \QSJ\ relate operator products of descendants to those of the primaries, while \SSJ\ constrain the terms that can appear on the RHS of the OPE of the primaries.

There are two more operators that annihilate $J(x)J(0)$,
\eqna{
\left[x^2Q_\alpha Q_\beta+Q_\alpha(ix\cdot\sigma\bar{S})_\beta-Q_\beta(ix\cdot\sigma\bar{S})_\alpha\right](J(x)J(0))&=0,\\
\left[x^2\bar{Q}_{\dot{\alpha}}\bar{Q}_{\dot{\beta}}+(S\,ix\cdot\bar{\sigma})_{\dot{\beta}}\bar{Q}_{\dot{\alpha}}-(S\,ix\cdot\bar{\sigma})_{\dot{\alpha}}\bar{Q}_{\dot{\beta}}\right](J(x)J(0))&=0,
}[JJannihilation]
thus constraining the OPE  $J(x)J(0)$. Other relations, giving OPEs of descendants in terms of the $J(x)J(0)$ primary OPE, are
\eqna{
j_\alpha(x)\bar{\jmath}_{\dot{\alpha}}(0) & =\frac{1}{x^4}\left[(S\,ix\cdot\sigma)_{\dot{\alpha}}(ix\cdot\sigma\bar{S})_\alpha-x^2\bar{Q}_{\dot{\alpha}}(ix\cdot\sigma\bar{S})_\alpha+2\Delta _Jx^2(ix\cdot\sigma)_{\alpha\dot{\alpha}}\right](J(x)J(0)),\displaybreak[0]\\
j_\mu(x)j_\nu(0) & =\frac{1}{16x^8}\left[(x^2\eta_{\mu\rho}-2x_\mu x_\rho)(S\sigma^\rho\bar{S}-\bar{S}\sigma^\rho S)\right.\displaybreak[0]\\
 & \quad\left.\quad\quad\quad\quad\quad\quad\quad\times\{x^4(\bar{Q}\bar{\sigma}_\nu Q-Q\sigma_\nu\bar{Q})+(x^2\eta_{\nu\lambda}-2x_\nu x_\lambda)(S\sigma^\lambda\bar{S}-\bar{S}\bar{\sigma}^\lambda S)\right.\displaybreak[0]\\
 & \quad\left.\quad\quad\quad\quad\quad\quad\quad\quad\quad\quad\quad\quad\quad\quad\quad\quad\quad\quad\quad-2x^2\left(Q\sigma_\nu\,ix\cdot\bar{\sigma}S-\bar{Q}\bar{\sigma}_\nu\,ix\cdot\sigma\bar{S}\right)\}\right.\displaybreak[0]\\
 & \quad\left.\quad\quad\quad-8i(\Delta_J+1)x^2(\eta_{\mu\nu}\eta_{\lambda\rho}-\eta_{\mu\lambda}\eta_{\nu\rho}-\eta_{\mu\rho}\eta_{\nu\lambda}-i\epsilon_{\mu\nu\lambda\rho})x^\lambda\right.\displaybreak[0]\\
 & \quad\left.\quad\quad\quad\quad\quad\quad\quad\times\{(x^2\eta^{\rho\delta}-2x^\rho x^\delta)S\sigma_\delta\bar{S}+x^2\bar{Q}\bar{\sigma}^\rho\,ix\cdot\sigma\bar{S}+4i\Delta_Jx^2x^\rho\}\right.\displaybreak[0]\\
 & \quad\left.\quad\quad\quad-8i(\Delta_J+1)x^2(\eta_{\mu\nu}\eta_{\lambda\rho}-\eta_{\mu\lambda}\eta_{\nu\rho}-\eta_{\mu\rho}\eta_{\nu\lambda}+i\epsilon_{\mu\nu\lambda\rho})x^\lambda\right.\displaybreak[0]\\
 & \quad\left.\quad\quad\quad\quad\quad\quad\quad\times\{(x^2\eta^{\rho\delta}-2x^\rho x^\delta)\bar{S}\bar{\sigma}_\delta S+x^2Q\sigma^\rho\,ix\cdot\bar{\sigma}S+4i\Delta_Jx^2x^\rho\}\right.\displaybreak[0]\\
 & \quad\left.\quad\quad\quad+32x^4\Delta_J(\Delta_J+1)(x^2\eta_{\mu\nu}-2x_\mu x_\nu)\right](J(x)J(0)),\displaybreak[0]\\
j_\mu(x)J(0) & =\frac{x^2\eta_{\mu\nu}-2x_\mu x_\nu}{4x^4}\left[S\sigma^\nu\bar{S}-\bar{S}\bar{\sigma}^\nu S\right](J(x)J(0)).
}
In sum, OPEs of the superdescendants are all determined from the primary OPE $J(x) J(0)$, and the superdescendants in $J(x)J(0)$ are constrained by superconformal symmetry and current conservation.  We will find the explicit expressions in the next subsection.

\subsec{Current-current OPEs using the superspace results of \texorpdfstring{\rcite{Osborn:1998qu}}{Osborn}}
We now consider  the superspace three-point functions \OsbornOOO\ where $\CO _1$ and $\CO_2$ are conserved currents, and for simplicity we take $\CO _1=\CO _2=\CJ$, so there is a $z_1\leftrightarrow z_2$ symmetry, implying the symmetry condition \tsymm\ on the function $t(X_3, \Theta _3, \bar \Theta _3)$ in \OsbornOOO.
The $\CJ$ superfield has the component expansion \Jzis.
We're interested in the three-point functions
\eqn{\vev{\CJ(z_1)\CJ(z_2) \CO ^{\mu _1\dots \mu _\ell}(z_3)}=\frac{1}{\osbx{\bar 13}{2}\osbx{\bar 3 1}{2}\osbx{\bar 23}{2}\osbx{\bar 3 2}{2}}t_{\CJ\CJ \CO _\ell }^{\mu _1\dots \mu _\ell}(X_3, \Theta _3, \bar \Theta _3).}[JJksope]
The scaling relation \tscaling, with $q=\bar q=1$ for the conserved currents has
$a=\tfrac{1}{3}(q_k+2\bar q_k)-2$ and $\bar a=\tfrac{1}{3}(\bar q_k+2q_k)-2.$  We now discuss the constraints on $t$ in \JJksope\ coming from current conservation.  The condition that $\CJ$ is conserved, written in superspace as $D^2\CJ=\bar D^2\CJ=0$, implies that ${\cal D}^2t=\bar {\cal D}^2t=0$, where ${\cal D}$ acts on $t$ as differential operators as in \dtops.

A first consequence is that  the operator  $\CO_3$ in \JJksope\ must be a real operator of vanishing R-charge and integer spin $\ell$ (much as in the $\Phi \bar \Phi$ OPE of the previous subsection).
  Suppose, to the contrary, that e.g.\ $R(\CO _k)=2$, which would lead to $\bar a=a+1$ in \tscaling, which would fix $t\stackrel{?}{=}\bar \Theta _3^2 f(X_3)$ ($f$ can't have any additional factors of $\bar \Theta_3$, since $\bar\Theta _3^{n>2}=0$, nor $\Theta _3$ factors without spoiling \tscaling). But that $t$ cannot satisfy $\bar{\cal D}^2t=0$.   One can similarly use ${\cal D}^2t=\bar{\cal D}^2t =0$ to exclude all other possibilities for non-zero R-charge operators in \JJksope.   So, in what follows, we take $\CO _\ell$ to have $q=\bar q =\half \Delta $, and thus $a=\bar a=\half \Delta -2$ in \tscaling.

The conditions ${\cal D}^2t=\bar{\cal D}^2t=0$ uniquely determine the function $t^{\mu _1\dots \mu _\ell}$ in \JJksope.  Let's first write it for spin-$\ell =0$ operators $\CO _k$ in \JJksope:
\eqn{t_{\CJ\CJ \CO _{\ell =0}}(X, \Theta, \bar \Theta)=\frac{c_{JJ\CO}}{(X \cdot \bar X)^{2-\half \Delta }}\left[1-\frac{1}{4}(\Delta -4)(\Delta -6)\frac{\Theta  ^2\bar\Theta ^2}{X\cdot \bar X}\right],}[JJspinz]
with $c_{JJ\CO_{\ell=0}}$ an arbitrary coefficient.  Because the coefficient of the term involving $\Theta _3$ and $\bar \Theta _3$ is determined, the superconformal descendant three-point functions are determined from that of the superconformal primaries.  The case \JJJsope\ where all three operators are conserved currents, $q_k=1$, is exceptional, since ${\cal D}^2 X^{-2}=0$ (up to contact terms).

For the case of an $\ell =1$ superconformal primary operator $\CO _k^\mu$ in \JJksope, the conditions determine, much as in \rrt
\eqn{t_{\CJ\CJ \CO_{\ell =1}}^{\mu}(X, \Theta, \bar \Theta)=\frac{c_{JJ\CO_{\ell=1}}}{(X \cdot \bar X)^{\frac{5}{2}-\half \Delta }}\left[X_-^\mu -\frac{ \Delta -5}{\Delta-2}\frac{(X_-\cdot X_+)X_+^\mu}{X \cdot \bar X}\right],}[JJspini]
where $X_+^\mu \equiv \half (X^\mu +\bar X^\mu)$, called $Q^\mu$ in \rcite{Osborn:1998qu}, is odd under the $z_1\leftrightarrow z_2$ operation in \tsymm, and $X_-^\mu \equiv i(X^\mu -\bar X^\mu)\equiv -4\Theta \sigma^\mu\bar \Theta$, called $P^\mu$ in \rcite{Osborn:1998qu}, is even under the $\mathbb{Z}_2$.  An example of a real, primary $\ell =1$ operator is the FZ operator $\CT _\mu$ \FZis, with $\Delta _{\CT _\mu}=3$.   If we set  $\Delta _{\CO ^\mu }=3$ in \JJspini\ and $\Delta _\CO=2$ in \rrt,  the two expressions properly coincide.

For general, even-spin-$\ell$ superconformal primary $\CO_k^{(\mu _1\dots \mu _\ell)}$, \JJspinz\ generalizes to
\eqn{t_{\CJ\CJ \CO_{\ell \text{ even}}}^{(\mu _1\dots \mu _\ell)} =c_{JJ\CO _{\ell}}\frac{X_+^{(\mu _1}\cdots X_+^{\mu _\ell)}}{(X \cdot \bar X)^{2- \frac12(\Delta -\ell)}}\left[1-\frac{1}{4}(\Delta-\ell-4)(\Delta+\ell-6)\frac{\Theta ^2\bar \Theta ^2}{X \cdot \bar X}\right]-\text{traces}.}[JJspine]
The generalization of \JJspini\ for odd spin $\ell$  is
\eqn{t_{\CJ\CJ \CO_{\ell \text{ odd}}}^{(\mu _1\dots \mu _\ell)} =c_{JJ\CO _{\ell}}\frac{X_+^{(\mu _1}\cdots X_+^{\mu _{\ell-1}}}{(X \cdot \bar X)^{2-\frac12(\Delta -\ell)}}\left[X_-^{\mu _\ell)} -\frac{\Delta -\ell-4}{\Delta -2}\frac{(X_-\cdot X_+)X_+^{\mu _\ell)}}{X \cdot \bar X}\right]-\text{traces}.}[JJspino]
In both \JJspine\ and \JJspino\ the $\ell$ Lorentz indices are symmetrized, with the traces removed, to obtain a spin-$\ell$ irreducible Lorentz representation.

These superspace results encode all component three-point functions, giving relations among the conformal primary components.    To make this explicit, we need to expand  both sides of \JJksope\ in the Grassmann coordinates; we expand $\CJ (z_1)$ and $\CJ (z_2)$ as in \Jzis, and  $\CO ^{\mu _1\dots \mu _\ell}$ is as in \realo, and likewise on the RHS.  Then, matching the coefficients of the terms with powers of the Grassmann coordinates  $\theta _{i=1,2,3}$ and $\bar\theta_{i=1,2,3}$ on the two sides of \JJksope, gives relations among the primary and descendant three-point functions analogous to \ADrelni\ and \PSDrelns.   For $\ell$ even, \JJspine\ gives a contribution when we take all three operators to be primary, setting all Grassmann coordinates to zero; the coefficient $c_{JJ\CO _\ell}$ of this primary contribution determines all descendant three-point function.  For $\ell$ odd, the three-point function with all three operators primary vanishes, as does \JJspino\ when all Grassmann coordinates are set to zero, but there are still non-zero superconformal descendant contributions and expanding \JJspino\ gives relations among them.

The three-point function result  \JJksope, with \JJspine\ and \JJspino, can be expanded in the Grassmann coordinates.  To illustrate this, let's now expand the  three-point function in $\theta _3\equiv\theta$ and $\bar \theta _3\equiv\bar{\theta}$, setting  $\theta _{1,2}=0$, and $\bar \theta _{1,2}=0$.   Using \Xexp\ we have
\eqna{X_+^\mu| _{\theta _{i=1,2}=\bar\theta _{i=1,2}=0}&=Z^\mu +{\color{red}2Y^{\mu \nu}\theta\sigma _\nu \bar \theta}+Z^2\left({\color{red}\frac{x_{12}^\mu}{r_{12}}}-Z^\mu\right)\theta^2\bar\theta^2,\cr
X_-^\mu| _{\theta _{i=1,2}=\bar\theta _{i=1,2}=0}&=-2(Z^2\eta ^{\mu \nu }-2Z^\mu Z^\nu)\theta\sigma_\nu \bar \theta,\cr
\left.\frac{\Theta ^2\bar\Theta ^2}{X\cdot \bar X}\right| _{\theta _{i=1,2}=\bar\theta _{i=1,2}=0}&=Z^2\theta^2\bar \theta^2.
}[Xpmexp]
One can also find
\eqna{X\cdot \bar X|_{\theta _{i=1,2}=\bar\theta _{i=1,2}=0}&=Z^2-2Z^4\left(2+{\color{red}\frac{x_{13}\cdot x_{23}}{r_{12}}}\right)\theta^2\bar{\theta}^2,\cr
X_+\cdot X_-|_{\theta _{i=1,2}=\bar\theta _{i=1,2}=0}&=2Z^2 Z^\mu \theta \sigma _\mu \bar \theta.}
So, for example, \JJspini\ becomes
\eqn{
\left.\frac{t^\mu_{\CJ \CJ \CO _{\ell =1}}}{\osbx{\bar 13}{2}\osbx{\bar 31}{2} \osbx{\bar 23}{2}\osbx{\bar 32}{2}}
\right|_{\theta _{i=1,2}=\bar\theta _{i=1,2}=0}=
-2\frac{C_{JJ\CO}}{r_{12}^2}Z^{\Delta -1}\left(Z^2\eta ^{\mu \nu}-\frac{\Delta+1}{\Delta-2}Z^\mu Z^\nu\right) \theta\sigma_\nu \bar\theta.
}
The red terms above drop out for primary correlation functions. Indeed, with no loss in generality, by using superconformal symmetry to map $(z_1, z_2, z_3)\to (0, x_2=\infty, z_3=z)$, the red terms map to zero, as discussed around \Thetamap.

For $\ell$ even, the results for $\vev{JJA^{\mu _1\dots \mu _\ell}}$,
$\vev{J J L^{\mu \mu _1\dots \mu _\ell}}$, and $\vev{JJ D^{\mu _1\dots \mu
_\ell}}$ coincide with those found in \rcite{Poland:2010wg} for the
corresponding quantities with $JJ$ replaced with $\phi \bar \phi$, while
$\vev{JJ M^{\mu \mu _1\dots \mu _\ell}}=0$ and $\vev{JJ N^{\mu _2\dots \mu
_\ell}}=0$.   Accounting for the distinction \rcite{Poland:2010wg} between
$L^{\mu \mu _1\dots \mu _\ell}$ and $L_{\prim}^{\mu \mu _1\dots \mu _\ell}$
and also between  $D^{\mu _1\dots \mu _\ell}$ and $D_{\prim} ^{\mu _1\dots
\mu _\ell}$, see \LDprim, we find, for $\ell $ even,\foot{We thank the
authors of \cite{Khandker:2014mpa} for
pointing out a mistake in our original formula for $\vev{JJD_\text{prim}}$.}
\eqna{\vev{JJ A^{\mu _1\dots \mu _{\ell}}}&=c_{JJ\CO _\ell}\frac{Z^{\Delta-\ell}}{r_{12}^2}Z^{\mu _1}\cdots Z^{\mu _\ell},\cr
\vev{JJ D_{\rm prim}^{\mu _1\dots \mu _\ell}}&=c_{JJ\CO
_\ell}\frac{(\Delta-2) (\Delta +\ell)(\Delta -\ell -2)}{8(\Delta -1)}\frac{Z^{\Delta+2-\ell}}{r_{12}^2}Z^{\mu _1}\cdots Z^{\mu _\ell}.\cr}[JJAD]
In addition, $\vev{JJM^{\mu \mu _1\dots \mu _\ell}_{\prim}}=0$ and $\vev{JJN^{\mu _2\dots \mu _\ell}_{\prim}}=0$, because the three-point function with $JJ$ can involve only even-spin operators.

Likewise, for $\ell $ odd, $\vev{JJ A^{\mu _1\dots \mu _\ell}}=0$ and
$\vev{JJD_\prim ^{\mu _1\dots \mu _\ell}}=0$, and the  non-zero primary
three-point functions are\foot{We thank the authors of ~\cite{Berkooz:2014yda} for pointing out a
mistake in our original
formula for $\vev{JJN_\text{prim}}$.}
\eqna{\vev{JJ M_{\rm prim}^{\mu \mu _1\dots \mu _\ell}}&=2 c_{JJ\CO
_\ell}\frac{ \Delta +\ell}{\Delta-2} \frac{Z^{\Delta -\ell}}{r_{12}^2}Z^\mu
Z^{\mu _1}\cdots Z^{\mu _\ell},\cr \vev{JJ N_{\rm prim}^{\mu _2\dots \mu
_\ell}}&=-2c_{JJ\CO
_\ell}\frac{(\ell+1)(\ell+2)(\Delta-\ell-2)}{\ell^2(\Delta-2)}\frac{Z^{\Delta
+2-\ell}}{r_{12}^2}Z^{\mu _2}\cdots Z^{\mu _\ell},\cr}[JJMN] with $\vev{JJ
A^{\mu _1\dots \mu _\ell}}=0$ and $\vev{JJD_\prim ^{\mu _1\dots \mu
_\ell}}=0$.   In all of the above it's to be understood that the $Z^\mu$'s
are symmetrized with the traces removed.  For all $\ell$,  $\vev{J
(x_1)J(x_2)L_\prim ^{\mu \mu _1\dots \mu _\ell}}=0$, because the primary
three-point function necessarily involves only the single coordinate
$Z^\mu$, and it is impossible to use that to build an operator with the
right Lorentz index structure to match $L_{\prim}^{\mu \mu _1\dots \mu
_\ell}$.

Summarizing, we find the relations
\eqna{c_{JJD_{\ell;\text{prim}}}&=-\frac{\Delta (\Delta +\ell)(\Delta -\ell -2)}{8(\Delta -1)}c_{JJ A_\ell},\cr
c_{JJN_{\ell-1}}&=-\frac{(\ell+1)(\ell+2)(\Delta-\ell-2)}{\ell^2 (\Delta+\ell)} c_{JJ M_{\ell +1}},\cr
c_{JJ L_{\text{\prim}}}&=0.}[JJrelns]
A check on these results is that $c_{JJD_{\ell;\text{prim}}}$ and $c_{JJN_\ell}$ properly vanish when $\CO^{\mu _1\dots \mu _\ell}$ saturates the unitarity bound, $\Delta =\ell +2$, as then the components $N_{\prim}$ and $D_{\prim}$ become null states and must vanish.  As a special case,  for $\ell =1$ and $\Delta _\CO =3$, we have $\CO ^{\mu}={\cal T}^\mu$, the Ferrara--Zumino supermultiplet, where $M^{\mu \nu}\sim T^{\mu \nu}$ and $N\sim T^{\phantom{\mu}\!\mu} _\mu =0$.

Upon going to components, the resulting two- and three-point functions can be converted to expressions for the OPE coefficients, including conformal descendants, as in \FCreln.  The superconformal descendant relations can then be determined by using the two-point and three-point function relations discussed in the previous paragraph.  A more efficient approach would be to convert directly in superspace, from the two-point and three-point function results above, to sOPE expressions.   A special case has been explicitly worked out in \rcite{Dolan:2000uw}, as outlined after \chiralring. For our case of interest here, i.e.\ two conserved currents,
\eqn{\CJ(z_1)\CJ(z_2)=\sum _{\substack{\rm{sprimary}\\ \CO_\Delta ^{(\ell )}}} c_{JJ}^{\CO}F _\Delta ^{(\ell)}(z_{12}, \partial _{x_2}, \partial _{\theta _2}, \partial _{\bar \theta _2}) \CO_\Delta ^{\ell} (z_2),}[JJFope]
with $F$ determined by requiring that using this and two-point functions \OsbornOO\ on the LHS of \JJksope\ reproduces the RHS of \JJksope. For example, for $\ell=0$, $F$ satisfies
\eqn{\frac{1}{\osbx{\bar{1} 3}{2}\osbx{\bar{3} 1}{2}\osbx{\bar{2} 3}{2}\osbx{\bar{3}2}{2}}t^{(\ell=0)}_{\CJ\CJ \CO}(X_3,\Theta_3,\bar{\Theta}_3)=c_{JJ\CO }F^{(\ell=0)}_\Delta (z_{12},\partial_{x_2},\partial_{\theta_2},\partial_{\bar{\theta}_2})\frac{1}{\osbx{\bar{2}3}{\Delta _\CO }\osbx{\bar{3}2}{\Delta _\CO}},}
where $t$ on the LHS is given in \JJspinz.

\newsec{Four-point function conformal blocks}[JJJJsec]

Four-point functions (more generally $n$-point functions) can be reduced and computed via the OPE.   For a four-point function $\vev{\CO _i(x_1)\CO _j (x_2) \CO _r(x_3)\CO _s(x_4)}$, one can apply the OPE \opegen\ to $\CO _i(x_1)\CO _j(x_2)$, and also to $\CO_r(x_3)\CO _s(x_4)$, reducing the four-point function to sums of two-point functions between the resultants on the RHS of the two OPE pairs:
\eqn{\vev{\CO _i(x_1) \CO _j (x_2) \CO _r (x_3)\CO _s(x_4)}=\sum _{\substack{\text{primary}\\ \CO _k}}\frac{1}{r_{12}^{\frac12(\Delta _i+\Delta _j)}r_{34}^{\frac12(\Delta _r+\Delta _s)}}c_{ijk} c^{k\ell}c_{\ell rs} g_{\Delta _k, \ell _k}(u, v),}[fourpoint]
where $u\equiv r_{12} r_{34}/r_{13} r_{24}\equiv z\bar z$ and $v\equiv r_{14} r_{23}/r_{13} r_{24}\equiv (1-z)(1-\bar z)$ are the two independent conformal cross-ratios for four-point functions.  The four-point function conformal blocks $g_{\Delta, \ell}$ are fixed functions \rcite{Dolan:2000ut,Dolan:2003hv} that account for the sum over descendants\foot{As in \rcite{Rattazzi:2010yc}, we find it convenient to modify the original definition of $g_{\Delta, \ell}$ by dropping a $(-\half )^\ell$ factor: $g_{\Delta , \ell}^{\text{here}}=(-2)^\ell g_{\Delta, \ell}^{\text{D\&O}}$.}
\eqna{g_{\Delta, \ell}(u, v)&=\frac{z\bar z}{z-\bar z}\left( k_{\Delta +\ell}(z)k_{\Delta -\ell -2}(\bar z)-(z\leftrightarrow \bar z)\right)\\ k_\beta (x)&\equiv x^{\beta /2}{}_2F_{1}(\beta /2, \beta/2, \beta ; x).}[gis]
The decomposition \fourpoint\ is in the $s$ channel of the four-point function, and one can of course alternatively compute in the $t$ channel or the $u$ channel, and all three must of course agree.  There is a recent and growing literature on exploring these crossing symmetry relation constraints, following \rcite{Rattazzi:2008pe}.

The fact that the sum in \fourpoint\ for  non-SUSY ${\cal N}=0$ theories can be reduced to a sum over primaries, with the descendant contributions accounted for in the universal conformal block functions $g_{\Delta , \ell}$, is a powerful consequence of the fact that conformal symmetry completely determines the descendant contributions to the OPE from those of the primaries.  As we have emphasized, the analogous statement generally does not hold for superconformal primaries.  So, in superconformal theories, there is generally no analog of \fourpoint\ involving only a sum over only the superconformal primaries.  In a nutshell, there is no universal notion of ``superconformal blocks" analogous to \gis.   One can define superconformal blocks for correlation functions of short multiplets, as we'll discuss and review, but they depend on the particular operators in the correlation function and are still not universal.

In this section, we will discuss the ${\cal N}=1$ conformal blocks for $\vev{JJJJ}$ and $\vev{JJ\phi \phi ^*}$.    These two cases are expected to be nicer than generic four-point functions in ${\cal N}=1$ SCFTs, because the operators are in shortened representations, and that determines the coefficients of all superconformal descendants in the intermediate channel in terms of those of the superconformal primaries.\foot{As we emphasized, that seems to not be the case for generic ${\cal N}=1$ operators, so it seems that generic four-point functions can not be reduced to a set of ${\cal N}=1$  superconformal blocks depending only on the superconformal primaries.}

\subsec{Review of the \texorpdfstring{${\cal N}=1$}{N=1} conformal blocks for \texorpdfstring{$\vev{\phi \phi ^*\phi \phi ^*}$}{<phi phi* phi phi*>} \texorpdfstring{\rcite{Poland:2010wg,Vichi:2011ux}}{Poland}}

The four point function of two chiral and two anti-chiral operators can be expanded as
\eqn{\langle \phi (x_1)\phi ^*(x_2)\phi (x_3)\phi ^*(x_4)\rangle =\frac{1}{r_{12}^{\Delta _\phi}r_{34}^{\Delta _\phi}}\sum _{\CO _\ell\in \phi \times\phi ^*} \frac{(c_{\phi \phi ^*A_\ell})^2}{c_{A_\ell A_\ell}}{\cal G}_{\Delta, \ell}^{\phi \phi ^*; \phi \phi ^*}(u, v)}
where ${\cal G}_{\Delta, \ell}^{\phi \phi ^*; \phi \phi ^*}(u, v)$ is a superconformal block that account for the $s$-channel OPE sum over the $A_\ell$, $M_{\ell +1}$, $N_{\ell-1}$, and $D_\ell$ conformal primaries, along with their descendants.  Using \PSDrelns\ and
\ADnorms, the result is \rcite{Poland:2010wg} (accounting for $g_{\Delta, \ell}^{\text{here}}=(-2)^\ell g_{\Delta, \ell}^{\text{D\&O}}$)
\eqna{
{\cal G}_{\Delta, \ell}^{\phi \phi ^*; \phi \phi ^*}=g_{\Delta, \ell}+\frac{\Delta +\ell}{4(\Delta +\ell +1)}g_{\Delta +1, \ell +1}&+\frac{\Delta -\ell -2}{4(\Delta -\ell -1)}g_{\Delta +1, \ell -1} \\
&\hspace{1cm}+\frac{(\Delta +\ell)(\Delta -\ell -2)}{16(\Delta +\ell +1)(\Delta -\ell -1)}g_{\Delta +2, \ell}.}[cacai]
As we have emphasized, there is not a general notion of superconformal block, and the superscript in ${\cal G}_{\Delta, \ell}^{\phi \phi ^*; \phi \phi ^*}$ emphasizes that this superconformal block applies only for this specific channel and four-point function.

Indeed, computing the same $\vev{\phi (x_1)\phi ^* (x_2) \phi (x_3) \phi ^* (x_4)}$ in the channel where the $x_1$ and $x_3$ operators are brought together, leads to an intermediate sum over very different classes of operators, corresponding to \chiralring.  We can define ${\cal G}^{\phi \phi ; \phi ^*\phi ^*}_{\Delta, \ell}$ for this class, and the result involves a single $g_{\Delta, \ell}$, rather than the four terms \cacai\ found in the $s$ channel.  See \rcite{Vichi:2011ux} for some of the details.  This illustrates that there isn't a universal notion of superconformal blocks, even for different channels of the same four-point function.

\subsec{The \texorpdfstring{${\cal N}=1$}{N=1} conformal blocks for \texorpdfstring{$\vev{JJJJ}$ and $\vev{JJ\phi \phi ^*}$}{<JJJJ>}}

The four-point current correlator can be expanded as
\eqn{\vev{J(x_1)J(x_2)J(x_3)J(x_4)}=\frac{1}{r_{12}^2r_{34}^2}\sum _{\CO _{\Delta, \ell}\in J\times J}\frac{(c_{JJA_\ell})^2}{c_{A_\ell A_\ell}} {\cal G}_{\Delta, \ell }^{JJ; JJ}(u, v),}
where the ${\cal N}=1$ superconformal blocks on the RHS account for the sum over the $A_\ell$, $M_{\ell +1}$, $N_{\ell -1}$, and $D_{\ell}$ primaries in the intermediate operators \realo, along with their descendants.   Comparing with \fourpoint, the decomposition in terms of ${\cal N}=0$ blocks simply follows from squaring the coefficients in \JJrelns\ and dividing by the normalizations in \ADnorms.   For $\ell$ even we find
\eqn{{\cal G}^{JJ; JJ}_{\Delta,\, \ell\text{ even}}=g_{\Delta ,
\ell}+\frac{(\Delta-2)^2(\Delta +\ell)(\Delta -\ell -2)}{16\Delta(\Delta +\ell +1)(\Delta -\ell -1)}g_{\Delta +2, \ell}.}[Jblocke]
 For $\ell$ odd we find (with here an arbitrary overall normalization choice)
\eqn{{\cal G}^{JJ; JJ}_{\Delta, \,\ell\text{ odd}}=g_{\Delta +1 , \ell
+1}+\frac{(\ell +2)^2(\Delta+\ell+1)(\Delta -\ell
-2)}{\ell^2(\Delta+\ell)(\Delta -\ell -1)} g_{\Delta +1, \ell-1}.}[Jblocko]

We can immediately now also obtain the conformal blocks for
\eqn{\vev{J(x_1)J(x_2)\phi (x_3)\phi ^*(x_4)}=\frac{1}{r_{12}r_{34}^{\Delta _\phi}}\sum _{\CO _{\Delta, \ell}}\frac{c_{JJ\CO _\ell}c_{\phi \phi ^* \CO _\ell}}{c_{\CO _\ell\CO _\ell}}{\cal G}_{\Delta, \ell }^{JJ; \phi \phi ^*}(u, v),}[JJca]
where
\eqn{{\cal G}_{\Delta, \,\ell \text{ even}}^{JJ; \phi \phi ^*}=g_{\Delta ,
\ell}+\frac{(\Delta-2)(\Delta +\ell)(\Delta -\ell -2)}{16\Delta(\Delta +\ell +1)(\Delta -\ell -1)}g_{\Delta +2, \ell},}[JJcae]
\eqn{{\cal G}_{\Delta,\, \ell \text{ odd}}^{JJ; \phi \phi ^*}=g_{\Delta +1
, \ell +1}-\frac{(\ell +2)(\Delta+\ell+1)(\Delta -\ell
-2)}{\ell(\Delta+\ell)(\Delta -\ell -1)} g_{\Delta +1, \ell-1}.}[JJcao]

\subsec{Connection with Dolan and Osborn's \texorpdfstring{${\cal N}=2$}{N=2} conformal blocks for \texorpdfstring{$\vev{\varphi \varphi \varphi \varphi}$}{<phi phi phi phi>} \texorpdfstring{\rcite{Dolan:2001tt}}{Dolan}}

In ${\cal N}=2$ SCFTs, operators are labeled by their $SU(2)_I$ representation $I=0, \half, \dots$, value of $I_3$, their $U(1)^{{\cal N}=2}$ charge, in addition to dimension $\Delta$ and spins $(j, \bar \jmath)$. Several ${\cal N}=1$ representations assemble together to form a single ${\cal N}=2$ superconformal representation.    The ${\cal N}=1$ $U(1)_R^{{\cal N}=1}$ is given by (see e.g.\ \rcite{Benini:2009mz})
\eqn{R^{{\cal N}=1}=\frac{1}{3}R^{{\cal N}=2}+\frac{4}{3}I_3.}[Nrcharges]
Taking the ${\cal N}=2$ supercharges $Q_\alpha ^I$ to have $R^{{\cal N}=2}$ charge $-1$, then $Q_\alpha ^{I=1,2}$ has $R^{{\cal N}=1}$ charges $1/3$ and $-1$, with the latter the ${\cal N}=1$ supercharge.

In particular, an ${\cal N}=2$ conserved current supermultiplet has primary components with $I=1$, $R^{{\cal N}=0}=0$, $\Delta =2$, $\ell =0$.  It
consists of an ${\cal N}=1$ conserved current supermultiplet $\CJ$, plus a ${\cal N}=1$ chiral multiplet and conjugate anti-chiral multiplet $\bar \Phi$, with $\Delta =2$, $\ell =0$.   The primary components were called $\varphi ^{ij}$ in  $\varphi ^{(ij)}$ of \rcite{Dolan:2001tt}, and we denote them as \eqn{\begin{pmatrix}\varphi ^{11}\cr \varphi ^{(12)}\cr \varphi ^{22}\end{pmatrix}= \begin{pmatrix}\phi \cr J\cr \phi ^*\end{pmatrix}=\begin{pmatrix} |I=1, I_3=1\rangle\cr |I=1, I_3=0\rangle\cr |I=1, I_3=-1\rangle\end{pmatrix}.}[niicompts]
The structure of the four-point function for this ${\cal N}=2$ supermultiplet was considered in \rcite{Dolan:2001tt}, and a variety of possible four-point function conformal blocks, corresponding to the possible intermediate operator in the OPE, were presented.  The recent work \rcite{Poland:2010wg} used these results to connect with the ${\cal N}=1$ superconformal blocks ${\cal G}^{\phi \phi ^*; \phi \phi ^*}$.   In this section, we connect the ${\cal N}=2$ results of \rcite{Dolan:2001tt} with our ${\cal N}=1$ results for ${\cal G}^{JJ;JJ}$ and ${\cal G}^{JJ; \phi \phi ^*}$.

The $SU(2)_I$ symmetry implies that when we take the $\varphi \varphi $ OPE we get representations ${\bf 3}\otimes {\bf 3}={\bf 1}\oplus {\bf 3}\oplus {\bf 5}$, i.e.\ the RHS can have
representations  $I=0,1,2$,  of $SU(2)_I$.  When we consider the $\vev{\varphi \varphi \varphi \varphi}$ four-point function, the contributions thus can be labeled by the $I=0,1,2$ values of the intermediate operators.  Following \rcite{Dolan:2001tt}, we refer to these contributions as $A_0$, $A_1$, and $A_2$, respectively.  The $SU(2)_I$ symmetry implies that the various four-point functions in $\vev{\varphi \varphi \varphi \varphi }$ are governed by the group theory of Clebsch--Gordan coefficients (following \rcite{Dolan:2001tt}, we absorb $A_0$'s Clebsch, $\frac{1}{3}$, into its normalization):
\eqna{{\cal G}^{{\cal N}=2|\phi \phi;  \phi ^*\phi ^*}&=A_2, \cr
{\cal G}^{{\cal N}=2| \phi \phi ^*; \phi \phi ^*}&=A_0 +\frac{1}{2} A_{1}+\frac{1}{6} A_{2},\cr
{\cal G}^{{\cal N}=2| JJ; JJ}&=A_0+\frac{2}{3} A_{2}, \cr
{\cal G}^{{\cal N}=2| JJ; \phi \phi ^*}&=A_0-\frac{1}{3}A_{2}.}[cgdecomp]

The functions $A_0$, $A_1$, and $A_2$ get independent contributions from each possible ${\cal N}=2$ superconformal multiplets that can appear in the intermediate channel of the $\varphi \varphi$ OPE.    Since the supercharges have $I=\half$, each contributing ${\cal N}=2$ superconformal multiplet has operators with different $I$ values, that can potentially contribute to all three $A_{I=0,1,2}$.    A variety of ${\cal N}=2$ supermultiplets and their $A_{0,1,2}$ contributions were presented in  \rcite{Dolan:2001tt}.   We will apply \cgdecomp\ to their results to determine the multiplet's contribution ${\cal G}^{{\cal N}=2|\phi \phi ; \phi ^*\phi ^*}$, ${\cal G}^{{\cal N}=2|\phi \phi ^*; \phi \phi ^*}$, ${\cal G}^{{\cal N}=2|JJ; JJ}$, and ${\cal G}^{{\cal N}=2|JJ; \phi \phi ^*}$.  Decomposing the ${\cal N}=2$ multiplet into multiplets under the ${\cal N}=1$ subalgebra, these ${\cal N}=2$ superconformal blocks decompose into sums of ${\cal N}=1$ superconformal blocks.  The case ${\cal G}^{{\cal N}=2|\phi \phi ^*; \phi \phi ^*}\to {\cal G}^{{\cal N}=1|\phi \phi^*; \phi \phi ^*}$ was presented in
 \rcite{Poland:2010wg}, and here we'll similarly discuss a few simple examples of \cgdecomp.

One class of examples are the shortened ${\cal N}=2$ multiplets containing at most twist $\Delta -\ell =2$ operators.  Quoting \rcite{Dolan:2001tt} (with $g_{\Delta, \ell}^{\text{D\&O}}=(-2)^{-\ell}g_{\Delta, \ell}^{\text{here}}$), these have
\eqn{\begin{gathered}
A_0=g_{\Delta =\ell +2, \ell}+\frac{(\ell +2)^2}{4(2\ell +3)(2\ell +5)}g_{\Delta = \ell +4, \ell +2},\cr
A_1=g_{\Delta = \ell +3, \ell +1}, \qquad A_2=0.
\end{gathered}}[DOAshortl]
An example in this class is the ${\cal N}=2$ conserved current multiplet \niicompts, which corresponds to setting $\ell =-1$ in \DOAshortl.  Another example in this class is the ${\cal N}=2$ stress-energy tensor multiplet, corresponding to $\ell =0$ in \DOAshortl; this ${\cal N}=2$ multiplet contains the ${\cal N}=1$ stress-tensor multiplet \FZis\ together with the ${\cal N}=1$ current multiplets of $SU(2)_I$.   We see from \cgdecomp\ that, since $A_2=0$, no operators in this class contribute to ${\cal G}^{{\cal N}=2|\phi \phi; \phi ^*\phi ^*}$.  Their contributions to ${\cal G}^{{\cal N}=2|\phi \phi ^*; \phi \phi ^*}$ fit with the decomposition of these ${\cal N}=2$ multiplets into ${\cal N}=1$ multiplets and the results of \rcite{Poland:2010wg}, as was presented there.   The blocks given in \Jblocke, \Jblocko, \JJcae, \JJcao\ for this case, $\Delta =\ell +2$, contain only a single ${\cal N}=0$ block, ${\cal G}^{JJ; JJ}_{\Delta=\ell +2, \ell}=g_{\Delta=\ell+2, \ell}$ and ${\cal G}^{JJ; \phi \phi ^*}_{\Delta =\ell +2, \ell}=g_{\Delta =\ell +2, \ell}$.  The result \cgdecomp\ and \DOAshortl\ for ${\cal G}^{{\cal N}=2|JJ; JJ}_{\Delta=\ell +2, \ell}$ and ${\cal G}^{{\cal N}=2|JJ; \phi \phi ^*}_{\Delta = \ell +2, \ell}$ contain contributions from two ${\cal N}=1$ real multiplets in the ${\cal N}=2$ multiplet, with primary components $\CO _{\Delta=\ell +2, \ell}$ and $\CO '_{\Delta=\ell +4, \ell +2}$, and  the relative coefficient in \DOAshortl\ accounts for the ${\cal N}=2$ relation among their OPE coefficients.

To quote a more complicated ${\cal N}=2$ representation multiplet, the contributions to the conformal blocks from the multiplet of operators and descendants when the primary has $R^{N=2}=0$, $I=0$, for general $\Delta $ and $\ell$, is \rcite{Dolan:2001tt}
\eqna{A_{0}(u, v)&=g_{\Delta , \ell}+\frac{1}{12}g_{\Delta +2, \ell}+\frac{(\Delta +\ell +2)^2}{16(\Delta +\ell +1)(\Delta +\ell +3)}g_{\Delta +2, \ell+2}\cr &\quad +\frac{(\Delta -\ell)^2}{16(\Delta -\ell -1)(\Delta -\ell+1)}g_{\Delta +2, \ell -2}\cr &\quad+\frac{(\Delta +\ell +2)^2(\Delta -\ell)^2}{256(\Delta +\ell +1)(\Delta +\ell +3)(\Delta -\ell -1)(\Delta -\ell +1)}g_{\Delta +4, \ell},\cr
A_{1}(u, v)&=g_{\Delta +1, \ell +1}+ g_{\Delta +1, \ell -1}+\frac{(\Delta +\ell +2)^2}{16(\Delta +\ell +1)(\Delta +\ell +3)}g_{\Delta +3, \ell +1}\cr&\quad+\frac{(\Delta -\ell)^2}{16(\Delta -\ell -1)(\Delta -\ell +1)}g_{\Delta +3, \ell -1},\cr
A_{2}(u, v)&=g_{\Delta +2, \ell}}[jrblocks]
Using \cgdecomp, we can read off the contributions to ${\cal G}^{{\cal N}=2}$ from this representation.  The case ${\cal G}^{{\cal N}=2| \phi \phi ^*; \phi \phi ^*}$ was considered in \rcite{Poland:2010wg} and decomposed there in terms of the ${\cal N}=1$ blocks.  The other cases in \cgdecomp\ can be similarly analyzed.

\newsec{Discussion \& Conclusion}[CONC]

The current-current (s)OPE $\CJ (z_1) \CJ (z_2)$ can have only real $R^{{\cal N}=1}=0$ operators of even spin $\ell$ and their descendants on the RHS.  For non-Abelian groups, odd-$\ell $ real operators can also contribute, proportional to the group's structure constants $f_{abc}$.   The constraints of ${\cal N}=1$ superconformal symmetry, combined with the current conservation, imply relations among the OPE coefficients, essentially giving the super-descendant coefficients in terms of those of the super-primaries.

We also gave the basic ${\cal N}=1$ superconformal blocks for ${\cal G}_{\Delta, \ell}^{JJ; JJ}$ and ${\cal G}_{\Delta, \ell}^{JJ; \phi \phi ^*}$.  These are analogous to the ${\cal G}_{\Delta, \ell}^{\phi \phi ^*; \phi \phi ^*}$ superconformal blocks given in \rcite{Poland:2010wg} and the ${\cal G}_{\Delta, \ell}^{\phi \phi; \phi ^*\phi ^*}$ described in \rcite{Vichi:2011ux}.  The blocks are analogous, but different, illustrating that there are no universal superconformal blocks.  In the ${\cal N}=2$ case, we discussed how these cases can be related using the $SU(2)_I$ Clebsch--Gordan coefficients  and the results of \rcite{Dolan:2001tt}.

We will explore some possible applications of the current-current OPE and superconformal symmetry to general gauge mediation of SUSY breaking in our upcoming paper \rcite{Fortin:2011aa}.

\ack{This research was supported in part by UCSD grant DOE-FG03-97ER40546. KI thanks the IHES for hospitality while some of this work was done, and the participants of the IHES Three String Generations and the CERN SUSY Breaking workshops for their comments. }

\appendix

\newsec{The (super)conformal algebra (and sign conventions)}[SCA]

This appendix both reviews standard material, and also attempts to give a consistent set of sign conventions.   The literature  contains many sign conventions (some with inconsistencies) for the conformal, supersymmetry, and superconformal algebras, so we will here elaborate a bit on our notation. (Our signs agree with e.g.\ \rcite{Minwalla:1997ka}.)  There are several places where sign errors can crop up.  One is a standard, but often obscured, sign difference when bosonic generators $A$ are replaced with differential operators $\CA$ acting on the coordinates,
\eqn{
[A, \CO]=-\mathcal{A}\CO.
}[bosdo]
  This is familiar from quantum mechanics, where $[H, \CO]=-i\hbar \partial_t \CO$, even though $H$ can be replaced with $\mathcal{H}=+i\hbar \partial _t$.   The sign in \bosdo\ accounts for the fact that transformations compose in the opposite order when acting on the coordinates.  Indeed, defining another transformation $[B, \CO]=-\CB\CO$, with $\CB$ the corresponding differential operator, the differential operators compose in the opposite order
\eqn{ AB(\CO)\equiv [A, [B, \CO]]=- [A, \CB \CO]=-\CB[A, \CO]=\CB \CA \CO.}[order]
So $[[A, B], \CO]=-[\CA , \CB]\CO$, which is consistent with \bosdo\ with $[A, B]=C$ and $[\CA, \CB]=\mathcal{C}$.  Many references, however,  do not make a notational distinction between what we're calling $A$ vs ${\cal A}$.   This issue is compounded in supersymmetry, see also
 \rcite{Dumitrescu:2011iu} for a very recent careful discussion.   As standard, we follow the conventions of  Wess \& Bagger \rcite{Wess:1992cp}.  The $Q$ analog of \bosdo\ in \rcite{Wess:1992cp} notation then has an $i$ but, potentially confusingly,  in \rcite{Wess:1992cp} no notational distinction is made between the operator vs the differential operator.  In addition, the metric of  \rcite{Wess:1992cp} is $\eta_{\mu\nu}=\diag(-1,1,1,1)$, with Hamiltonian $H=P^0=-P_0$, so now $[P_0, \CO]=+i\hbar \partial _0 \CO$ and ${\cal P}_0=-i\hbar \partial _0$.   We'll elaborate on these and related points  in what follows.

Recall (see e.g.\ \rcite{Osborn:1993cr}) that conformal transformations $x_\mu \to x'_\mu =(gx)_\mu $ are such that $dx'_\mu dx'{}^\mu =\Omega ^g(x)^{-2}dx_\mu dx^\mu$.  Beyond translations and rotations, this includes dilatations $x_\mu '=\lambda x_\mu$, with $\Omega ^g(x)=\lambda ^{-1}$, and special conformal transformations, $x_\mu '=(x_\mu -b_\mu x^2)/\Omega ^g(x)$, with $\Omega ^g=1-2b\cdot x + b^2 x^2$.   An operator is called (quasi-)primary if it transforms under all conformal transformations as $\CO(x)\to T(g)\CO (x)$, where
\eqn{(T(g)\CO)^i (x')=\Omega ^g(x) ^{\Delta _{\CO}}D^i_j\left({\cal R}^g_{\mu \alpha}=\Omega ^g\frac{\partial x'_\mu}{\partial x^\alpha}\right)\CO ^j(x),}[primt]
where $i$ labels the operator's representation $D^i_j$ of the Lorentz group, and $\Delta _{\CO}$ is the operator's scaling dimension.   For rotations and boosts, $\Omega ^g(x)=1$ and \primt\ is the standard Lorentz transformation of operators, with ${\cal R}^g_{\mu \alpha}$ the rotation or boost.  For dilatations, ${\cal R}^g_{\mu \alpha}=\delta _{\mu \alpha}$, so $D^i_j$ is the identity, and \primt\ is the standard scaling of operators with their scaling dimension $\Delta _{\cal O}$.  For
special conformal transformations only,  \primt\ proves restrictive, distinguishing the primary operators from the descendants.

On the LHS of \primt\ we've transformed both the operator and the coordinate, but we should replace $x\to g^{-1}x$ on both sides of \primt\ to get how the transformation acts on just the operator.  For example, the Poincar\'{e} generators act on the coordinates as $x^\mu \to x'{}^\mu = g(x^\mu)$,  and act on operators as  $g: \CO ^i(x)\to \CO '{}^i(x)=(U(g)\CO(x)U(g)^{-1})^i=D^i_j(g)\CO(g^{-1}(x))$, with $U(g)$ the appropriate unitary transformation.   Under general translations of operators forward by $a^\mu$,  via opposite action on the coordinates, $g_{a}: x^\mu \to x^\mu-a^\mu$, then $g_{a}: \CO (x)\to \CO'  (x)=U(a)\CO(x)U(a)^{-1}=\CO(x^\mu+a^\mu)$, with $U(a)=e^{-iP_\mu a^\mu}$.   We then have $[P_\mu ,\CO(x)]= i\partial _\mu \CO(x)$.  So the differential operator, as in \bosdo, is ${\cal P}^\mu = -i\partial _\mu$.  The minus sign in \bosdo\ and order reversal in \order\ are related to the $g^{-1}$ action on the coordinates.

 The dilatation generator acts on the coordinates as $g_\delta: x^\mu \to e^{-\delta}x^\mu$, $\Omega ^{g_\delta}(x)=e^\delta$, so $g_\delta:\CO(x)\to U(\delta )\CO(x)U(\delta )^{-1}=e^{\Delta _\CO\delta}\CO (e^\delta x)$, where $U(\delta)=e^{i\delta D}$.  This implies $[D, \CO]=-i(\Delta _\CO+x\cdot \partial)\CO$; hence the differential operator is ${\cal D}=i(\Delta _\CO+x\cdot \partial )$.      Now $g_\delta g_a: \CO(x)\to \CO (x+a)\to e^{\Delta _\CO\delta} \CO(e^\delta (x+a))=g_{e^{\delta }a}g_\delta\CO(x)$,  so $U(\delta) U(a)=U(e^\delta a)U(\epsilon)$, which implies $[P_\mu, D]=iP_\mu$.  The differential operators indeed correspondingly satisfy $[{\cal P}_\mu, {\cal D}]=i{\cal P}_\mu$.   We can likewise
take $U(b)=e^{-iK_\mu b^\mu}$ to generate special conformal transformations, and consider $U(b)U(\star)$ vs $U(\star)U(b)$ to get $[K_\mu , \star]$.  In sum, this yields the conformal group, that's isomorphic to $SO(d,2)$ in $d$ dimensions:
\begin{gather*}
[M_{\mu\nu},P_\rho]=i(\eta_{\mu\rho}P_\nu-\eta_{\nu\rho}P_\mu),\quad\quad [M_{\mu\nu},K_\rho]=i(\eta_{\mu\rho}K_\nu-\eta_{\nu\rho}K_\mu),\displaybreak[0]\\
[M_{\mu\nu},M_{\rho\sigma}]=i(\eta_{\mu\rho}M_{\nu\sigma}-\eta_{\nu\rho}M_{\mu\sigma}+\eta_{\nu\sigma}M_{\mu\rho}-\eta_{\mu\sigma}M_{\nu\rho}),\displaybreak[0]\\
[D,P_\mu]=-iP_\mu,\quad\quad [D,K_\mu]=iK_\mu,\quad\quad [K_\mu,P_\nu]=2i(\eta_{\mu\nu}D-M_{\mu\nu}),
\end{gather*}
where  $M_{\mu\nu}$ are the $SO(d-1,1)$ Lorentz generators.  Commutators not given are zero.

On a quasi-primary multi-component field $\mathcal{O}_I(x)$ we have
\eqn{
\begin{gathered}
\phantom{}[P_\mu, \mathcal{O}_I(x)] = i\partial_\mu \mathcal{O}_I(x),\quad\quad [D,\mathcal{O}_I(x)]=- i(x\cdot\partial+\Delta _\CO )\mathcal{O}_I(x),\\
[M_{\mu\nu},\mathcal{O}_I(x)]= i(x_\mu\partial_\nu-x_\nu\partial_\mu)\mathcal{O}_I(x)-\mathcal{O}_J(x)(s_{\mu\nu})^J_{\!\phantom{J}I},\\
[K_\mu,\mathcal{O}_I(x)]=i(x^2\partial_\mu-2x_\mu\,x\cdot \partial-2\Delta _\CO x_\mu)\mathcal{O}_I(x) +2(s_{\mu\nu})^J_{\!\phantom{J}I}x^\nu \mathcal{O}_J,
\end{gathered}
}[cdo]
where $(s_{\mu\nu})^J_{\!\phantom{J}I}$ are the appropriate finite-dimensional spin matrices obeying the  $M_{\mu\nu}$ algebra.

As an illustration of the order reversal in \order, consider  $[K_\nu, [P_\mu, \CO (0)]]$ and compare that to $[P_\mu, [K_\nu, \CO(0)]$ on a scalar primary operator at the origin.  The latter vanishes, since $K_\nu$ annihilates the scalar primary at the origin.  That is compatible with $[P_\mu, [K_\nu, \CO(0)]={\cal K}_\nu {\cal P}_\mu \CO (0)$ and ${\cal P}_\mu = -i\partial _\mu$ and ${\cal K}_\nu=-i(x^2 \partial _\nu -2x_\nu x\cdot \partial -2\Delta _\CO x_\nu)$.   The opposite order properly gives a non-zero result,  $[K_\nu, [P_\mu , \CO]]|_{x=0}= {\cal P}_\mu {\cal K}_\nu\CO|_{x=0}=2\Delta_\CO \eta _{\mu \nu}\CO (0)=-2i\eta _{\mu \nu}{\cal D}\CO |_{x=0}.$

We define the supersymmetry fermionic variations of operators as
\eqn{\delta_\xi \CO=i[\xi Q+\bar \xi\bar Q , \CO]\ =(\xi {\cal Q}+\bar \xi \bar {\cal Q})\CO,}[Qact]
where the $i$ after the first equality insures that, if $\CO$ is real, then so is $\delta _\xi\CO$.\foot{Recall  $(\xi Q)^\dagger = \bar \xi \bar Q$ in \rcite{Wess:1992cp} notation, where $\xi Q\equiv \xi ^\alpha Q_\alpha =-\xi_\alpha Q^\alpha= Q\xi$, and $\bar \xi \bar Q\equiv \bar \xi _{\dot \alpha} \bar Q^{\dot \alpha}=-\bar \xi ^{\dot \alpha}\bar Q_{\dot \alpha}=\bar{Q}\bar{\xi}$.}  In the second equality that $i$ is absent, and we use the superspace differential operators of \rcite{Wess:1992cp},
\eqn{{\cal Q}_\alpha=\frac{\partial}{\partial \theta ^\alpha}-i\sigma _{\alpha \dot \alpha}^\mu \bar \theta ^{\dot \alpha}\partial _\mu \quad \text{and}\quad \bar {\cal Q}_{\dot \alpha}=-\frac{\partial}{\partial \bar \theta ^{\dot \alpha}}+i\theta ^\alpha \sigma _{\alpha \dot \alpha}^\mu\partial _\mu.}[Qdo]
As in \order, the differential operators compose in the opposite order.  Note that
\eqn{\{{\cal Q}_\alpha, \bar{\cal Q}_{\dot \alpha}\}=2i\sigma _{\alpha\dot \alpha}^\mu \partial _\mu = -2\sigma _{\alpha\dot \alpha}^\mu \cal{P}_\mu;}[QPalg]
the last sign looks off, but it'll be OK, since \Qact\ gives (with $[\kappa Q, \bar \xi \bar Q]=\kappa ^\alpha \bar{\xi} ^{\dot \alpha} \{Q_\alpha, \bar{Q}_{\dot \alpha}\}$)
\eqn{(\delta_\kappa \delta _\xi -\delta _\xi \delta _\kappa)\CO =-(\kappa ^\alpha \bar \xi ^{\dot \alpha}-\xi ^\alpha \bar \kappa ^{\dot \alpha})[\{Q_\alpha, \bar{Q}_{\dot{\alpha}}\}, \CO] = -(\kappa ^\alpha \bar \xi ^{\dot \alpha}-\xi ^\alpha \bar \kappa ^{\dot \alpha})2 \sigma ^\mu _{\alpha \dot \alpha} [P_\mu, \CO],}
and also
\eqn{
(\delta_\kappa \delta _\xi -\delta _\xi \delta _\kappa)\CO =-(\kappa ^\alpha \bar \xi ^{\dot \alpha}-\xi ^\alpha \bar \kappa ^{\dot \alpha})\{{\cal Q}_\alpha, \bar {\cal Q}_{\dot \alpha}\}\CO= (\kappa ^\alpha \bar \xi ^{\dot \alpha}-\xi ^\alpha \bar \kappa ^{\dot \alpha})2 \sigma ^\mu _{\alpha \dot \alpha} {\cal P}_\mu \CO,
}
consistent with $[P_\mu, \CO]=-{\cal P}_\mu \CO$.  In short, if we use the notation of \rcite{Wess:1992cp} for the fermionic generators, the analog of \bosdo\ is
\eqn{Q(\CO)\equiv [Q, \CO\}=-i{\cal Q}\CO.}[fermi]
For a chiral superfield, $\Phi =\phi+\sqrt{2}\theta \psi+\cdots$, with $\bar Q_{\dot \alpha}(\Phi)=0$, we have $Q_\alpha (\phi )=-i\sqrt{2}\psi _\alpha$ etc.  For a real superfield $\CJ = J+i\theta j-i\bar \theta\bar \jmath+\cdots$, we find e.g.\ $Q_\alpha (J)=j_\alpha$ and $\bar Q_{\dot \alpha}(J)=-\bar \jmath_{\dot \alpha}$.

The superconformal algebra includes the usual supercharges $Q_\alpha$, $\bar{Q}_{\dot{\alpha}}$, superconformal supercharges, $S^\alpha$, $\bar{S}^{\dot{\alpha}}$, and the $U(1)_R$-current generator $R$.  The superconformal algebra includes, in addition to the conformal-algebra commutators,
\begin{gather*}
\{Q_{\alpha},\bar{Q}_{\dot{\alpha}}\}=2\sigma^\mu_{\alpha\dot{\alpha}}P_\mu,\quad\quad \{\bar{S}^{\dot{\alpha}},S^\alpha\}=2\bar{\sigma}^{\mu\dot{\alpha}\alpha}K_\mu,\displaybreak[0]\\
\{Q_\alpha,S^\beta\}=-i(\sigma^\mu\bar{\sigma}^\nu)_\alpha^{\!\phantom{\alpha}\beta}M_{\mu\nu}+\delta_{\alpha}^{\!\phantom{\alpha}\beta}(2iD+3R),\displaybreak[0]\\
\{\bar{S}^{\dot{\alpha}},\bar{Q}_{\dot{\beta}}\}=-i(\bar{\sigma}^\mu\sigma^\nu)^{\dot{\alpha}}_{\!\phantom{\dot{\alpha}}\dot{\beta}}M_{\mu\nu}-\delta^{\dot{\alpha}}_{\!\phantom{\dot{\alpha}}\dot{\beta}}(2iD-3R),\displaybreak[0]\\
[D,Q_\alpha]=-\tfrac{1}{2}iQ_\alpha,\quad\quad [D,\bar{Q}_{\dot{\alpha}}]=-\tfrac{1}{2}i\bar{Q}_{\dot{\alpha}},\quad\quad [D,S^\alpha]=\tfrac{1}{2}iS^\alpha,\quad\quad [D,\bar{S}^{\dot{\alpha}}]=\tfrac{1}{2}i\bar{S}^{\dot{\alpha}},\displaybreak[0]\\
[R,Q_\alpha]=-Q_\alpha,\quad\quad [R,\bar{Q}_{\dot{\alpha}}]=\bar{Q}_{\dot{\alpha}},\quad\quad [R,S^\alpha]=S^\alpha,\quad\quad [R,\bar{S}^{\dot{\alpha}}]=-\bar{S}^{\dot{\alpha}},\displaybreak[0]\\
[K^\mu,Q_\alpha]=-\sigma^\mu_{\alpha\dot{\alpha}}\bar{S}^{\dot{\alpha}},\quad\quad [K^\mu,\bar{Q}_{\dot{\alpha}}]=\sigma^\mu_{\alpha\dot{\alpha}}S^\alpha,\displaybreak[0]\\
[P^\mu,\bar{S}^{\dot{\alpha}}]=-\bar{\sigma}^{\mu\dot{\alpha}\alpha}Q_\alpha,\quad\quad [P^\mu,S^\alpha]=\bar{\sigma}^{\mu\dot{\alpha}\alpha}\bar{Q}_{\dot{\alpha}},\displaybreak[0]\\
[M_{\mu\nu},Q_\alpha]=-i\sigma_{\mu\nu\alpha}^{\!\phantom{\mu\nu\alpha}\beta}Q_\beta,\quad\quad [M_{\mu\nu},\bar{Q}_{\dot{\alpha}}]=i\bar{\sigma}_{\mu\nu\!\!\phantom{\dot{\beta}}\dot{\alpha}}^{\!\phantom{\mu\nu}\dot{\beta}}\bar{Q}_{\dot{\beta}},\displaybreak[0]\\
[M_{\mu\nu},S_\alpha]=-i\sigma_{\mu\nu\alpha}^{\!\phantom{\mu\nu\alpha}\beta}S_\beta,\quad\quad [M_{\mu\nu},\bar{S}_{\dot{\alpha}}]=i\bar{\sigma}_{\mu\nu\!\!\phantom{\dot{\beta}}\dot{\alpha}}^{\!\phantom{\mu\nu}\dot{\beta}}\bar{S}_{\dot{\beta}}.
\end{gather*}

The action of the superconformal generators on superfields was given in  \cite{Osborn:1998qu} in a very efficient and compressed notation, so we'll unpack it a bit here,  and write the variations as differential operators acting on superspace, with the $-1$ of \bosdo\ for the bosonic generators and the $-i$ of \fermi\ for the fermionic generators.   The ${\cal P}_\mu$, ${\cal Q}_\alpha$, and $\bar{\cal Q}_{\dot \alpha}$ are as given in \Qdo\ and \QPalg.  The ${\cal D}$ and ${\cal K}_\mu$ operators include Grassmann additions to the expressions found from \cdo, e.g.\ dilatations act as
$g_\delta : \CO (x, \theta, \bar\theta )\to e ^{i\delta D}\CO (x, \theta, \bar\theta)e^{-i\delta D}=e^{\Delta _\CO}\CO (e^\delta x, e^{\delta/2}\theta, e^{\delta /2}\bar \theta)$, which gives $[D, \CO]=-{\cal D}\CO$ with \eqn{
{\cal D}=i\left[x\cdot \partial +\half \left(\theta \frac{\partial}{\partial \theta}+\bar{\theta}\frac{\partial}{\partial \bar \theta}\right)+ \Delta\right].
}
For a $U(1)_R$ transformation, $g_R: \CO (x, \theta, \bar \theta)\to e^{i\alpha R} \CO (x, \theta, \bar\theta)e^{-i\alpha R}=e^{i\alpha r_\CO}\CO(x, e^{-i\alpha}\theta, e^{i\alpha}\bar\theta)$, so $[R, \CO]=-{\cal R}\CO$ with
\eqn{
{\cal R}=-r_\CO +\theta \frac{\partial}{\partial\theta} -\bar\theta\frac{\partial}{\partial{\bar\theta}}.
}

Finally, the special superconformal generators act on superfields as in \Qact,
\eqn{\delta _\eta\CO=i[\eta S+\bar \eta \bar S, \CO]=(\eta {\cal S}+\bar \eta \bar {\cal S})\CO,}[Sdo]
with ${\cal S}^\alpha$ and ${\bar {\cal S}}^{\dot \alpha}$ the differential operators acting on superspace, and we read off the transformation from that given in  \cite{Osborn:1998qu}: in the notation there
\eqn{\delta \CO ^i (z) =-{\cal L}\CO^i (z)+[\hat \omega _\alpha ^{\phantom{\alpha}\!\beta} (z_+)(s_\beta ^{\phantom{\beta}\!\alpha} )^i_{\phantom{i}\!i'}+\hat {\bar \omega}^{\dot \alpha}_{\phantom{\dot{\alpha}}\!\dot \beta}(z_-) (\bar s^{\dot \beta}_{\phantom{\dot{\beta}}\!\dot \alpha})^i_{\phantom{i}\! i'}]\CO ^{i'}(z)-2q\sigma (z_+)\CO ^i(z)-2\bar q \bar \sigma (z_-)\CO ^i (z),}
where ${\cal L}= (v^\mu (z_+)-2i\lambda (z_+)\sigma ^\mu \bar \theta) \partial _\mu +\lambda ^\alpha (z_+) D_\alpha +\bar \lambda _{\dot \alpha} (z_-)\bar D^{\dot \alpha},$
and $s$ and $\bar{s}$ act, respectively, on dotted and undotted indices, and form, respectively, spin-$j$ and spin-$\bar\jmath$ representations of the algebra.  Setting to zero the parameters for other transformations, we have
\begin{gather*}
v^\mu =-2\theta \sigma ^\mu \tilde{{\rm x}}_+\eta,\\
\lambda ^\alpha = -i(\bar \eta \tilde{{\rm x}}_+)^\alpha +2\eta ^\alpha \theta ^2, \qquad \bar{\lambda}^{\dot{\alpha}}=i(\tilde{{\rm x}}_+\eta)^{\dot{\alpha}}+2\bar{\eta}^{\dot{\alpha}}\bar{\theta}^2,\\
 \hat{\omega} _\alpha^{\phantom{\alpha}\!\beta} = 4\eta _\alpha \theta ^\beta +2\delta _\alpha ^{\phantom{\alpha}\!\beta} \theta \eta, \qquad \hat{\bar \omega}^{\dot \alpha}_{\phantom{\dot{\alpha}}\!\dot \beta}=-4\bar\theta ^{\dot \alpha}\bar \eta _{\dot \beta}-2\delta ^{\dot \alpha}_{\phantom{\dot{\alpha}}\!\dot \beta}\bar \eta \bar \theta,\\ \sigma =2\theta \eta, \qquad \bar \sigma =2\bar \eta \bar \theta,
 \end{gather*}
 where $\tilde{{\rm x}}_+=\tilde{{\rm x}}+2i\bar{\theta}\theta$. In our conventions we then find
 \eqn{{\cal S}^\alpha = i x\cdot \bar\sigma^{\dot \alpha \alpha}\bar {\cal Q}_{\dot \alpha}+2\theta^\alpha\left(\bar{\theta}\frac{\partial}{\partial\bar{\theta}}+\Delta+\frac32 r\right)+2\theta^\beta s_{\beta}^{\phantom{\beta}\!\alpha}+\theta^2\epsilon^{\alpha\beta}\left({\cal Q}_\beta+\frac{\partial}{\partial\theta^\beta}\right).}[sdo]

\bibliography{p1ref}


\end{document}